\documentclass[aps,10pt,twocolumn,superscriptaddress]{revtex4-2}

\newcommand{\externalizetikz}{0}   

\usepackage[english]{babel}

\usepackage[utf8]{inputenc}
\usepackage{mymath}
\usepackage{notation}
\usepackage{mathdots}
\usepackage{rotating}
\usepackage{nicefrac}

\usepackage{graphicx}
\usepackage[caption=false]{subfig}
\usepackage{makecell}
\usepackage{afterpage}
\usepackage{float}

\usepackage{xcolor}
\usepackage[color]{xy}

\usepackage{braket}
\usepackage{qcircuit}
\usepackage{environ}
\newcommand{\myqctmp}[2][0.4]{\Qcircuit @C=#2em @R=#1em @!R}
\newcommand{\colwidth}{0.75}
\NewEnviron{myqcircuit}[1][0.4]{\vcenter{\myqctmp[#1]{\colwidth} {\BODY}}}
\NewEnviron{myqcircuit*}[2]{\vcenter{\myqctmp[#1]{#2} {\BODY}}}
\renewcommand{\meter}{*=<1em,.9em>{\xy ="j","j"-<.778ex,.322ex>;{"j"+<.778ex,-.322ex> \ellipse ur,_{}},"j"-<0ex,.4ex>;p+<.5ex,.9ex> **\dir{-},"j"+<2.2ex,2.2ex>*{},"j"-<2.2ex,2.2ex>*{} \endxy} \POS ="i","i"+UR;"i"+UL **\dir{-};"i"+DL **\dir{-};"i"+DR **\dir{-};"i"+UR **\dir{-},"i" \qw}
\newcommand{\cgate}[2]{*+<.6em>[F*:#2]{#1} \POS ="i","i"+UR;"i"+UL **\dir{-};"i"+DL **\dir{-};"i"+DR **\dir{-};"i"+UR **\dir{-},"i" \qw}
\makeatletter
\newcommand{\multicgate}[3]{%
*+<1em,.9em>[F*:#3]{\hphantom{#2}}
\POS [#1,0] *+<1em,.9em>[F*:#3]{\hphantom{#2}}
\POS [0,0]="i",[0,0].[#1,0]="e",!C *+<1em,\dimexpr(\H@max)*#1+(\xymatrixrowsep@)*(#1-2)\relax>[F*:#3]{\hphantom{#2}}
\POS [0,0]="i",[0,0].[#1,0]="e",!C *+<1em,0em>[F*:#3]{#2},"e"+UR;"e"+UL **\dir{-};"e"+DL **\dir{-};"e"+DR **\dir{-};"e"+UR **\dir{-},"i" \qw}
\makeatother
\renewcommand{\ustick}[1]{*!D!<0em,-.3em>=<0em>{#1}}

\definecolor{myblue}{rgb}{0,0.4470,0.7410}
\definecolor{myred}{rgb}{0.8500,0.3250,0.0980}
\definecolor{myorange}{rgb}{0.9290,0.6940,0.1250}
\definecolor{mypurple}{rgb}{0.4940,0.1840,0.5560}
\definecolor{mygreen}{rgb}{0.4660,0.6740,0.1880}
\definecolor{mylightblue}{rgb}{0.3010,0.7450,0.9330}
\definecolor{mydarkred}{rgb}{0.6350,0.0780,0.1840}
\definecolor{mygrey}{rgb}{0.6, 0.6, 0.6}
\colorlet{Deltacolor}{mygreen!60!white}
\colorlet{Gammacolor}{Deltacolor!40!white}
\colorlet{Rzcolor}{Gammacolor!40!white}
\colorlet{CLkcolor}{myorange!60!white}
\colorlet{ECcolor}{mydarkred!60!white}
\colorlet{DickeOnecolor}{myblue!40!white}
\colorlet{DickeOneDcolor}{DickeOnecolor}
\colorlet{DickeOneUcolor}{DickeOnecolor}
\colorlet{DickeOneUDcolor}{DickeOnecolor}
\colorlet{DickeTwokNcolor}{DickeOnecolor}
\colorlet{DickeTwokNUcolor}{DickeTwokNcolor}
\colorlet{DickeTwokNDcolor}{DickeTwokNcolor}
\colorlet{DickeTwokNUDcolor}{DickeTwokNcolor}
\definecolor{PRcolor}{rgb}{0.6,0.6,0.6}
\definecolor{PLcolor}{rgb}{0.6,0.6,0.6}
\definecolor{subPRcolor}{rgb}{0.85,0.85,0.85}
\definecolor{mutedgreen}{rgb}{0.6,0.85,0.6}

\ifnum\externalizetikz=2
  \usepackage{tikzexternal}
  \tikzsetexternalprefix{fig/}
\else
  \usepackage{tikz}
  \usepackage{pgfplots}
  \usetikzlibrary{calc}
  \usetikzlibrary{positioning}
  \usetikzlibrary{plotmarks}
  \pgfplotsset{
    compat=newest,
    table/header=false,
    tick label style={font=\footnotesize},
    label style={font=\small},
    legend style={font=\footnotesize},
    legend cell align=left,
    colormap={parula}{
      rgb255=(53,42,135)
      rgb255=(15,92,221)
      rgb255=(18,125,216)
      rgb255=(7,156,207)
      rgb255=(21,177,180)
      rgb255=(89,189,140)
      rgb255=(165,190,107)
      rgb255=(225,185,82)
      rgb255=(252,206,46)
      rgb255=(249,251,14)
    }
  }
  \pgfplotsset{
    myColOne/.style={myblue},
    myColTwo/.style={myred},
    myColThr/.style={myorange},
    myColFou/.style={mypurple},
    myColFiv/.style={mygreen},
    myColSix/.style={mylightblue},
    myColSev/.style={mydarkred}
  }
  \ifnum\externalizetikz=1
    \usepgfplotslibrary{external}
  \fi
\fi

\usepackage{tabularx,multirow,booktabs}
\newcolumntype{L}{>{\raggedright\arraybackslash}X}
\newcolumntype{C}{>{\centering\arraybackslash}X}
\newcolumntype{R}{>{\raggedleft\arraybackslash}X}

\usepackage[colorlinks,urlcolor=blue,linkcolor=black,citecolor=blue]{hyperref}
\usepackage[capitalize,nameinlink]{cleveref}
\usepackage{orcidlink}

\begin{document}

\title{Efficient LCU block encodings through Dicke states preparation}

\author{Filippo Della~Chiara\orcidlink{0009-0001-7966-439X}}
\email{filippo.dellachiara3@studio.unibo.it}
\affiliation{These authors contributed equally to this work.}
\affiliation{Dipartimento di Matematica,
Università di Bologna, 40127 Bologna, Italy}

\author{Martina Nibbi\orcidlink{0009-0001-6440-0498}}
\email{martina.nibbi@tum.de}
\affiliation{These authors contributed equally to this work.}
\affiliation{Technical University of Munich, School of Computation, Information and Technology, Boltzmannstra{\ss}e 3, 85748 Garching, Germany}

\author{Yizhi Shen\orcidlink{0000-0002-4160-5482}}
\email{yizhis@lbl.gov}
\affiliation{Applied Mathematics and Computational Research Division,
Lawrence Berkeley National Laboratory, Berkeley, CA 94720, USA}

\author{Roel Van~Beeumen\orcidlink{0000-0003-2276-1153}}
\email{rvanbeeumen@lbl.gov}
\affiliation{Applied Mathematics and Computational Research Division,
Lawrence Berkeley National Laboratory, Berkeley, CA 94720, USA}

\preprint{$\dagger$These authors contributed equally to this work.}

\begin{abstract}
With the Quantum Singular Value Transformation (QSVT) emerging as a unifying framework for diverse quantum speedups, the efficient construction of block encodings—their fundamental input model—has become increasingly crucial. However, devising explicit block encoding circuits remains a significant challenge.
A widely adopted strategy is the Linear Combination of Unitaries (LCU) method. While general, its practical utility is often limited by substantial gate overhead.
To address this, we introduce the \ourmethod{} (\cmlcu{}), a compact LCU formulation that requires only a linear number of ancilla qubits and is explicitly decomposed into one- and two-qubit gates. By exploiting the underlying Hamiltonian structure, we design a parametrized family of efficient Dicke state preparation routines, enabling systematic realization of the state preparation oracle at substantially reduced gate cost. The check matrix formalism further yields a constant-depth \select{} oracle, implemented as two fully parallelizable layers of singly controlled Pauli gates.
We construct explicit block encoding circuits for representative spin models such as the Heisenberg and spin glass Hamiltonians and provide detailed, non-asymptotic gate counts. Our numerical benchmarks confirm the efficiency of the \cmlcu{} approach, illustrating over an order-of-magnitude reduction in \cnot{} count compared to conventional LCU. This framework opens a pathway toward practical, low-depth block encodings for a broad class of structured matrices beyond those considered here.
\end{abstract}

\maketitle

\section{Introduction}

Quantum computing holds the promise of delivering meaningful speedups for a wide variety of physically and mathematically fundamental problems, including many-body simulation~\cite{Lloyd1996,Kempe2006,Lanyon2010,Babbush2023,Somma2025}, physical state preparation~\cite{Babbush18,Stanisic2022,chen2023,rouze2024}, and quantum linear algebra \cite{Harrow2009,Childs2017}. These tasks rely on central algorithmic primitives such as quantum phase estimation and real-time evolution.

In particular, the Quantum Singular Value Transformation (QSVT) provides a unifying framework for such algorithms \cite{Gilyen2018, Martyn2021, Motlagh2024}. QSVT enables the application of any polynomial transformation to the singular values of a target matrix by embedding it into a larger unitary,
a procedure known as block encoding. This embedding can be achieved using ancillary qubits and post-selection.
Despite its flexibility, the actual performance of QSVT-based algorithms is often constrained by the cost of the block encoding itself. In many cases, this overhead arises from the inability of generic block encoding techniques to effectively recognize and exploit the structure of the matrix to be embedded. Consequently, the corresponding quantum circuits can be resource-intensive
even on fault-tolerant devices.

Block encoding, therefore, constitutes a crucial component of quantum algorithm design, yet it is often treated as a black box. Only a handful number of works address its explicit construction \cite{Dong2021,Camps2022,Liu2025}, including notable exceptions tailored to sparse matrices \cite{Camps2024, Snderhauf2024,kuklinski2024}, hierarchical matrices \cite{Nguyen2022}, matrix product operators (MPOs) \cite{Nibbi_2024, Termanova_2024}, and variational circuits \cite{Kikuchi2023,Rullkotter2025}.

Among existing approaches, the Linear Combination of Unitaries (LCU) decomposition of the target matrix is the most widely employed technique for block encoding \cite{Childs2012, Gilyen2018, Chakraborty2024, Babbush_2019, Kane2025}. LCU represents a matrix as a weighted sum of unitaries and
realizes its block encoding via two oracles: a state-preparation routine
on the ancilla register and a sequence of multi-controlled unitaries, known as the \select{} oracle.
Although LCU is general and powerful, its textbook implementation can lead to a prohibitive gate count, primarily due to the use of multi-controlled operations in the \select{} oracle. 
Exploiting problem-specific structure can significantly reduce this gate complexity, as demonstrated in applications such as chemical Hamiltonians \cite{Babbush2018, Lee2021, loaiza2024, Wan2021, Boyd2024, Georges2025}, spin models \cite{Childs2018}, bosonic ladder operators \cite{simon2025}, and combinatorial optimization \cite{sanders2020}.

In this work, we present \emph{practical}, \emph{low-depth} LCU circuits for spin models, explicitly decomposed into one- and two-qubit gates.
By leveraging the structure of the underlying Hamiltonians, we significantly reduce the overall gate complexity of the block encoding circuits.
Our approach is built on two key ingredients.
First, we employ the preparation of \emph{Dicke states} \cite{Dicke_1954}, a special family of quantum states with a fixed number of excitations.
While balanced and unbalanced single-excitation Dicke states have been extensively studied and implemented in the literature \cite{Bartschi_2019, Bartschi_2022, Piroli2024, yu2024, Farrell2025}, we extend this framework by developing efficient constructions for double-excitation Dicke states subject to specific neighboring constraints.
Second, we exploit the \emph{check matrix} formalism \cite{Nielsen_2011, Georges2025}, enabling the  \select{} oracle to be implemented with only $\n$ \cnot{} and $\n$ \cz{} gates at the cost of a linear number of ancillae.
In near-term to early fault-tolerant regimes, where circuit depth remains a more pressing limitation than qubit count, this trade-off can contribute to a substantial practical advantage.
We refer to our approach as \emph{\ourmethod{}} (\cmlcu{}, pronounced “fox-LCU”).

We demonstrate the \cmlcu{}
block encoding on representative one-dimensional spin
Hamiltonians. Our circuits exhibit favorable gate-complexity scaling: linear \cnot{} scaling for the Heisenberg model and quadratic for the spin glass model. In simulations of systems with up to 16 spin sites, this yields more than an order of magnitude reduction in total \cnot{} gates compared to conventional LCU.
We further outline generalizations that extend the \cmlcu{} approach to broader classes of many-body Hamiltonians. Overall, these results underscore the practical benefits of embedding physical and/or algebraic structures into block encoding designs, offering a scalable path toward efficient quantum simulation and computation on near-term hardware.

All quantum circuit implementations of our \cmlcu{} block encodings are publicly available at \url{https://github.com/QuantumComputingLab/foqcs-lcu}.

\subsection{Road map}
\label{section:roadmap}
The paper is organized as follows. \cref{section:lcu} briefly reviews the standard LCU approach for block encoding non-unitary matrices. \cref{section:main_result} introduces the check matrix formalism and reveals how it leads to our \cmlcu{} construction. We start with the motivating single-qubit case in \cref{subsection: simple example} and proceed to the multi-qubit case in \cref{subsection: general implementation}, which yields a compact circuit representation. \cref{section:dicke} develops the relevant Dicke state preparation schemes as primitive subroutines. \cref{section:applications} presents two illustrative applications: block encodings of one-dimensional Heisenberg and spin glass Hamiltonians in \cref{section:heisenberg}
and \cref{section:spin_glass} respectively. \cref{section:generalization} generalizes the framework to broader classes of matrices. In \cref{section:resource estimation}, we provide end-to-end resource estimates, where we numerically demonstrate the efficiency of our algorithm in \cref{section:comparison}. We conclude and summarize in \cref{section:conclusion}.

\subsection{Notations and conventions}
We use Dirac notation, where $\bra{\cdot}$ and $\ket{\cdot}$ denote row and column vectors, respectively. The state $\ket{\basiselement}$ corresponds to the $\basiselement$-th element of the computational basis indexed by $\basiselement \in \{0, \ldots, 2^{\n}-1 \}$. Throughout this work, we adopt the following convention to represent $b$ in terms of its binary decomposition:
\begin{align}
\label{eq:binary_notation}
\begin{split}
\basiselement &= \left[\basiselementbit{\n-1}\cdots\basiselementbit{1}\basiselementbit{0}\right], \\
&= \basiselementbit{\n-1}\cdot2^{\n-1} + \cdots + \basiselementbit{1}\cdot 2^{1} + \basiselementbit{0}\cdot 2^{0},
\end{split}
\end{align}
with $\basiselementbit{\elle}\in \{0,1\}$ for $\elle=0,\ldots,\n-1$.
In particular, the state $\ket{2^\elle}$  corresponds to the bitstring containing a single $1$ at position $\basiselementbit{\elle}$ and $0$ elsewhere:
\begin{equation}
\ket{2^\elle} = \ket{\underbrace{0\cdots 0}_{\n-1-\elle} 1 \underbrace{0\cdots 0}_{\elle}}.
\end{equation}
Following the same logic, the state $\ket{2^\elle+2^k}$ has only two $1$s at positions $\basiselementbit{\elle}$ and $\basiselementbit{k}$.

Given a generic operator $U$, we next define $U^{\basiselementbit{\elle}}$ as:
\begin{equation}
    U^{\basiselementbit{\elle}} =
    \begin{cases}
        \Ip & \text{if $\basiselementbit{\elle}=0$} \\
        U & \text{if $\basiselementbit{\elle}=1$}
    \end{cases}.
\end{equation}
This is equivalent to controlling $U$ by the $\ell$-th qubit:
\begin{equation}\label{eq:def_controlled_matrices}
\begin{myqcircuit}[0]
& \lstick{\ket{\basiselementbit{\n-1}}} & \qw & \qw \\
& \lstick{\raisebox{1ex}{$\vdots~~$}} \\
& \lstick{\ket{\basiselementbit{\elle}}} & \ctrl{4} & \qw \\
& \lstick{\raisebox{1ex}{$\vdots~~$}} \\
& \lstick{\ket{\basiselementbit{1}}} & \qw & \qw \\
& \lstick{\ket{\basiselementbit{0}}} & \qw & \qw \\
& \lstick{} & \gate{U} & \qw
\end{myqcircuit}
\end{equation}

\begin{figure*}
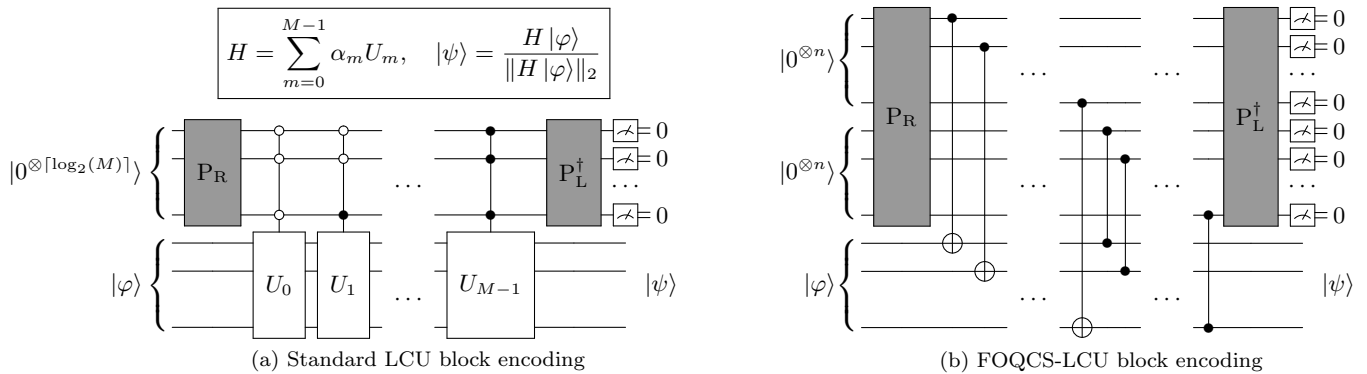

    \centering
    \qquad\quad\ %
    \subfloat[Standard LCU block encoding\label{fig:lcu_general}]{%
    \begin{minipage}{\columnwidth}
    \centering
    \fbox{$
    \Ham = \Sum_{\m=0}^{\M-1}\pscoeff_\m \Umat_\m, \quad
    \ket\psi = \Frac{\Ham\ket\varphi}{\normtwo{\Ham\ket\varphi}}
    $} \\[1.75ex]
    \input{figures/lcu_general}
    \end{minipage}
    }%
    \hfill%
    \subfloat[\cmlcu{} block encoding\label{fig:be_general}]{%
    \input{figures/be_general}
    }%
    \label{fig:lcu_be_general}
    \caption{(a) Standard LCU and (b) \cmlcu{} block encodings circuits.}
\end{figure*}

\section{Linear combination of unitaries}
\label{section:lcu}

One of the most widely used block encoding techniques directly exploits the Linear Combination of Unitaries (LCU) decomposition of the target Hamiltonian \cite{Childs2012}:
\begin{equation}
    \Ham = \sum_{\m=0}^{\M-1}\pscoeff_\m \Umat_\m,
\end{equation}
where the scalar coefficients $\pscoeff_m$ are generally complex.
At worst, the number of terms $\M$ can scale exponentially with the system size $n$. However, this decomposition admits polynomial or even linear scaling in many physically relevant settings, including condensed matter physics and quantum chemistry.
For this reason, LCU is still considered the state-of-the-art approach for block encodings in the field of quantum simulation~\cite{Babbush_2019, Lee2021, Loaiza_2023, Sze_2025}.

LCU relies on a state preparation oracle pair $(\PR,\PL)$ and a $\select$ oracle \cite{Gilyen2018, Childs2012, Chakraborty2024}.
As shown in \cref{fig:lcu_general}, the oracles $\PR$ and $\PL$ only act on the $\lceil\log_2(\M)\rceil$ ancillae and prepare, respectively, the following states:
\begin{align}
    \label{eq:def_PR}
    \ket{0^{\otimes\lceil\log_2(\M)\rceil}} & \xrightarrow{\PR} \frac{1}{\sqrt{\generalnorm}} \sum_{\m=0}^{\M-1}\sqrt{\alpha_m} \ket{\m}, \\
    \label{eq:def_PL}
    \ket{0^{\otimes\lceil\log_2(\M)\rceil}} & \xrightarrow{\PL} \frac{1}{\sqrt{\generalnorm}} 
 \sum_{\m=0}^{\M-1}\conjugate{\sqrt{\alpha_m}}\ket{\m},
\end{align}
where $\ket{\m}$ denotes the computational basis state indexed by the binary representation of the integer $\m$ while $\generalnorm$ is the normalization constant:
\begin{equation}\label{eq:def_normalization_lcu}
    \generalnorm = \normone{\alpha} = \sum_{\m=0}^{\M-1} \abs{\alpha_m}.
\end{equation}
This corresponds to rescaling the Hamiltonian spectrum by a factor of $\generalnorm$.

The $\select$ oracle, on the other hand, applies $\M$ multi-controlled unitaries acting on the system register, with each unitary controlled by the associated ancillary state $\ket{\m}$:
\begin{equation}
    \select = \sum_{\m=0}^{\M-1} \ket{\m}\bra{\m}\otimes \Umat_m.
\end{equation}

Note that the circuit in \cref{fig:lcu_general} correctly encodes $\Ham/\generalnorm$ when the ancillae are all measured in the $\ket{0}$ state. The corresponding success probability of post-selection is \cite{Nibbi_2024}:
\begin{equation}
    \prob_{\text{success}} = \sum_{\lambda}\left\vert\frac{\lambda}{\generalnorm}\right\vert^2 \cdot \left\vert \braket{\lambda|\varphi}\right\vert^2,
\end{equation}
where $\lambda$ and $\ket{\lambda}$ denote the eigenvalues and eigenvectors of the target matrix $\Ham$, respectively, and $\ket{\varphi}$ is the initial system state.

For the standard LCU block encoding, the dominant resource overhead arises from implementing the multi-controlled gates in the \select{} oracle \cite{Mikko_2004,Low_2019, Nibbi_2024, Wan2021}.
Nonetheless, this overhead can be substantially reduced by introducing extra ancillae and exploiting the structure of the problem at hand. Notably, \cite{Babbush2018,Lee2021} achieved linear T-gate complexity for fermionic operators, with similar improvements reported in \cite{simon2025} for ladder operators.

\section{\ourmethod}
\label{section:main_result}

We now present a practical and efficient LCU implementation based on the check matrix formalism \cite{Nielsen_2011}, using Pauli strings as the underlying unitaries. 
Since the Pauli matrices form a complete basis for the space of $2 \times 2$ complex matrices, a decomposition always exists.
The key feature of this LCU formulation is that the \select{} oracle requires no multi-controlled unitaries. We refer to our approach as the \emph{\ourmethod{}} (\cmlcu{}), pronounced as ``fox-LCU''.

In \cref{subsection: simple example}, we present a basic example to illustrate how our implementation works. In \cref{subsection: general implementation}, we generalize this approach to matrices of dimension $2^\n \times 2^\n$.

\subsection{\cmlcu{} for a $2\times2$ matrix}
\label{subsection: simple example}
We start with the single-qubit case by observing that a complex matrix $\generalmat \in \mathbb{C}^{2 \times 2}$ can be written as:
\begin{align}
    \label{eq:lcu_simple_1}
    \generalmat &= \pscoeff_{00} \Ip + \pscoeff_{01}\Zp + \pscoeff_{10}\Xp + \pscoeff_{11}\Yp, \\
    \label{eq:lcu_simple_2}
    &= \pscoeff_{00} \Ip + \pscoeff_{01}\Zp + \pscoeff_{10}\Xp -i\pscoeff_{11}\Zp\Xp,
\end{align}
where we assume $\normone{\pscoeff} = 1$ and use the identity $\Yp = -i\Zp\Xp$.
This notation recalls the check matrix formalism in \cref{table:check_matrix_formalism} \cite{Nielsen_2011}, so that the coefficient $\pscoeff_{\ixbit{}\jzbit{}}$ corresponds to the pair $(\ixbit{},\jzbit{})$.

\begin{table}
\begin{tabular}{c|c} 
    \text{Pauli} & $(\ixbit{},\jzbit{})$ \\
    \hline
    $\Ip$ & $(0,0)$ \\
    $\Zp$ & $(0,1)$ \\
    $\Xp$ & $(1,0)$ \\
    $\Yp$ & $(1,1)$
\end{tabular}
\caption{Check matrix formalism for the Pauli matrices.}
\label{table:check_matrix_formalism}
\end{table}

By using a state preparation pair $(\PR,\PL)$, each of which prepares the following state on two ancilla qubits:
\begin{alignat}{3}
\ket{00}&\xrightarrow{\PR}\, && \sqrt{\pscoeff_{00}}\ket{00}+\sqrt{\pscoeff_{01}}\ket{01}+ \nonumber\\
&&&\sqrt{\pscoeff_{10}}\ket{10}+\sqrt{-i\pscoeff_{11}}\ket{11}, \label{eq:def_prep_1q_PR} \\[0.5em]
\nonumber \ket{00}& \xrightarrow{\PL}\, && \conjugate{\sqrt{\pscoeff_{00}}}\ket{00}+\conjugate{\sqrt{\pscoeff_{01}}}\ket{01}+ \nonumber\\
&&&\conjugate{\sqrt{\pscoeff_{10}}}\ket{10} +\conjugate{\sqrt{-i\pscoeff_{11}}}\ket{11}, \label{eq:def_prep_1q_PL}
\end{alignat}
we realize an LCU block encoding for $\generalmat$ that requires no multi-controlled operations in the \select{} oracle. The $\Xp$ and $\Zp$ contributions are activated by the first ($\ixbit{}$) and second ($\jzbit{}$) ancilla, respectively:
\begin{equation}
\begin{myqcircuit}
\lstick{\ket0}    & \multicgate{1}{\PR}{PRcolor}    & \ctrl{2} & \qw
                  & \multicgate{1}{\PLdag}{PLcolor} & \meter   & \rstick{0}\cw \\
\lstick{\ket0}    & \ghost{\PR}           & \qw      & \ctrl{1}
                  & \ghost{\PLdag}        & \meter   & \rstick{0}\cw \\
\lstick{\ket\varphi} & \qw                   & \targ    & \control\qw
                  & \qw                   &
  \rstick{\frac{\generalmat\ket\varphi}{\norm{\generalmat\ket\varphi}}}\qw
\end{myqcircuit}
\label{eq:be_general_simple}
\end{equation}
An explicit proof of this simple case can be found in \cref{sec:proof-1q}.

\subsection{\cmlcu{} for an $\n$-qubit matrix}
\label{subsection: general implementation}
The circuit in \cref{eq:be_general_simple} can be generalized to $\n$ qubits, as shown in \cref{fig:be_general}, and is applicable to arbitrary linear combinations of Pauli strings.
Given the standard Hamiltonian decomposition into Pauli strings:
\begin{equation}\label{eq:def_lcu_pauli}
    \Ham = \sum_{\m=0}^{\M-1} \pscoeff_\m \bigotimes_{\elle=0}^{\n-1}\sigma_{\m_\elle},
\end{equation}
where $\sigma_{\m_\elle}\in\{\Ip,\Xp,\Yp,\Zp\}$ and $\Vert\pscoeff\Vert_1=1$, we can then map every $\sigma_{\m_\elle}$ to a binary pair $(\ixbit{\elle}, \jzbit{\elle})$ using the check matrix formalism in \cref{table:check_matrix_formalism}.
As a result, we have the identity:
\begin{equation}
    \sigma_{\m_\elle} = (-i)^{\ixbit{\elle}\cdot\jzbit{\elle}}\, \Zp^{\jzbit{\elle}}\Xp^{\ixbit{\elle}},
\end{equation}
where $\Xp^{\ixbit{\elle}}$ and $\Zp^{\jzbit{\elle}}$ follow the definition from \cref{eq:def_controlled_matrices}.

Next, we combine the single-qubit indices $\ixbit{\elle}$ and $\jzbit{\elle}$ for $\elle=0,\ldots,\n-1$ to get the binary representations of two $n$-qubit indices $\ix$ and $\jz$ as in \cref{eq:binary_notation}:
\begin{align}
    \ix &= [\ixbit{\n-1}\cdots\ixbit{1}\ixbit{0}], &
    \jz &= [\jzbit{\n-1}\cdots\jzbit{1}\jzbit{0}],
\end{align}
so that every $\m$ in \cref{eq:def_lcu_pauli} corresponds to some tuple of integers $(\ix,\jz)$.

Finally, by expanding the summation over every $\ix,\jz\in\{0,\ldots,2^\n-1\}$, with $\pscoeff_{\ix\jz}=0$ if there is no corresponding $\pscoeff_\m$ in the original summation, and by absorbing the complex coefficients:
\begin{equation}\label{eq:lcu_newpscoeff}
        \newpscoeff_{\ix\jz} = 
            (-i)^{ \sum_\elle \ixbit{\elle} \cdot \jzbit{\elle}}\pscoeff_{\ix\jz},
\end{equation}
we obtain the \emph{check matrix decomposition} of $\Ham$:
\begin{equation}\label{eq:lcu_general_check_matrix}
    \Ham = \sum_{\ix=0}^{2^{\n-1}} \sum_{\jz=0}^{2^{\n-1}} \newpscoeff_{\ix\jz} \bigotimes_{\elle=0}^{{\n-1}}\Zp^{\jzbit{\elle}}\Xp^{\ixbit{\elle}}.
\end{equation}
Note that even if the summation formally extends over $4^\n\geq \M$ terms, the normalization constraint in \cref{eq:def_normalization_lcu} remains unchanged as the new coefficients $\newpscoeff_{\ix\jz}$ are either equal to the old ones up to a complex phase or identically zero.

Starting from \cref{eq:lcu_general_check_matrix}, we can then define the action of \PR{} for the $\n$-qubit case:
\begin{equation}
\label{eq:PR_state_general}
    \PR\ket{0^{\otimes\n}}\ket{0^{\otimes\n}} = \sum_{\ix=0}^{2^{\n-1}} \sum_{\jz=0}^{2^{\n-1}} \sqrt{\newpscoeff_{\ix\jz}} \ket{\ix}\ket{\jz},
\end{equation}
while $\PL$ prepares the complex conjugate:
\begin{equation}
\label{eq:PL_state_general}
    \PL\ket{0^{\otimes\n}}\ket{0^{\otimes\n}} = \sum_{\ix=0}^{2^{\n-1}} \sum_{\jz=0}^{2^{\n-1}} \conjugate{\sqrt{\newpscoeff_{\ix\jz}}} \ket{\ix}\ket{\jz}.
\end{equation}
The $\select$ oracle, as illustrated in \cref{fig:be_general}, can be realized by only $\n$ $\cnot$ and $\n$ $\cz$, each forming a fully parallelizable layer.
The proof of correctness for the general $n$-qubit case can be found in \cref{sec:proof-nq}.

This $\select$ oracle implementation eliminates the need for multi-controlled gates, has constant circuit depth equal to $2$, and is independent of the specific matrix being encoded.
We note that \cite{Georges2025} discusses a comparable Pauli decomposition strategy in the context of first-quantized molecular Hamiltonians, though our approach differs fundamentally in its circuit design.
These techniques stand in sharp contrast to the standard LCU construction, where the $\select$ oracle is typically the most resource-intensive component due to the need to decompose the multi-controlled operations into elementary gates.

In the \cmlcu{} approach, all information associated with the encoded matrix is instead fully shifted to the state preparation stage.
More specifically, the \PR{} and \PL{} oracles always need $2\n$ ancillae, compared to the $\lceil\log_2(\M)\rceil$ ancillae for the standard LCU.
In the worst-case scenario where $\M = 4^n$, the ancilla requirements of the two approaches, as well as the computational cost of \PR{} and \PL{}, become equal.
If, on the other hand, $\M$ is small, this also reflects on the sparsity of the new coefficients $\newpscoeff$.
A natural first simplification of \PR{} and \PL{} may then come from employing an algorithm for the preparation of sparse quantum states, with complexities ranging from $\bigO{\M\n}$ to $\bigO{\M\n/\log{n}}$ if we allow for extra ancillaries \cite{Malvetti2021,Zhang2022,Li2025}.
However, if the problem exhibits additional structure, this structure can be exploited to further reduce the cost of the state preparation. This approach is demonstrated in \cref{section:applications} for the Heisenberg and spin glass Hamiltonians, where we achieve preparation complexities of $\bigO{M}$.

Overall, the \cmlcu{} features a remarkably simple $\select$ oracle and, when paired with structure-aware state preparation techniques, enables a significantly more efficient block encoding compared to the standard LCU implementation. These results point to its potential applicability across a broader class of problems.

\section{Preparation of Dicke states}
\label{section:dicke}

Before discussing practical applications of the \cmlcu{} block encoding, we introduce several subcircuits that will be used in the following sections. In particular, we summarize key results from the literature on the preparation of both \emph{balanced} and \emph{unbalanced} Dicke states \cite{Dicke_1954, Bartschi_2019, Farrell2025}, denoted by $\ket{\Dicke{\n}{\excitations}}$ and $\ket{\Dicke{\n}{\excitations}(\alpha)}$, respectively. These Dicke states are quantum states of $\n$ qubits with exactly $\excitations$ excitations.
We also introduce efficient circuits for new categories of Dicke states 
that will play a critical role in implementing the $\PR$ and $\PL$ oracles for the spin Hamiltonians considered in this paper.

In \cref{section:dicke_one_balanced}, we review the definition of balanced Dicke states and present the implementation for the case with a single excitation from \cite{Bartschi_2019}. 
In \cref{section:dicke_one_unbalanced}, we show how to generalize the previously defined subcircuit to handle the unbalanced case.
In \cref{section:dicke_two_kN}, we define a new subclass of Dicke states with two excitations and subject to neighboring constraints. 
Finally, in \cref{section:dicke_double}, we introduce \emph{double Dicke states} where the same Dicke state is efficiently duplicated on two separate ancillae registers.

\subsection{Balanced Dicke states with $\excitations=1$ excitation}
\label{section:dicke_one_balanced}
The balanced Dicke state $\ket{\Dicke{\n}{1}}$, also known as the $W$-state, is defined as:
\begin{align}
\label{eq:dicke one def}
    \ket{\Dicke{\n}{1}} &= \frac{1}{\sqrt{\n}}\left(\ket{\cdots001}+\ket{\cdots010}+\cdots+\ket{100\cdots}\right), \nonumber\\
    &= \frac{1}{\sqrt{\n}} \sum^{\n-1}_{\elle=0} \ket{2^\elle}.
\end{align}
This state with a single excitation can be prepared in a linear number of \cnots{}, as shown in~\cite{Bartschi_2019}. 

To construct the corresponding circuit, we rely on two auxiliary components: the $\subr{\theta}$ and $\staircase{\n}$ subcircuits.
They are depicted in \cref{fig:subr_gamma} and \cref{fig:staircase}, respectively, where we also highlight the recursive structure of $\staircase{\n}$. 
More specifically, the angles $\theta_\elle$ used in \cref{fig:staircase} are given by:
\begin{align} \label{eq:def_theta_dicke1}
    \theta_\elle &= 2\arccos\sqrt{\frac{1}{\elle+1}}, & \elle&\in\{1,2,\ldots,\n-1\}.
\end{align}

The balanced Dicke state $\ket{\DickeOne{\n}}$ can then be prepared using the following circuit \cite{Bartschi_2019}:
\begin{equation}
\label{eq:def_dicke1_prep}
    \begin{myqcircuit}
        \lstick{\ket{0^{\otimes\n}}} & \cgate{\DickeOne{\n}}{DickeOnecolor} & \rstick{\ket{\DickeOne{\n}}}\qw
    \end{myqcircuit}
\end{equation}
where:
\begin{equation}\label{eq_oracle_dicke_one}
\begin{myqcircuit}
& {/\strut^{\n}}\qw & \cgate{\DickeOne{\n}}{DickeOnecolor} &\qw \\
\end{myqcircuit}
\;=\;
    \begin{myqcircuit}
& {/\strut^{\n-1}}\qw & \multicgate{1}{\staircase{\n}}{Deltacolor} & \qw \\
&\gate{\Xp}& \ghost{\staircase{\n}} & \qw  
    \end{myqcircuit}
\end{equation}
It is important to note that the Pauli $\Xp$ gate plays a key role in activating the Dicke state preparation, since $\staircase{\n}\ket{0^{\otimes \n}} = \ket{0^{\otimes \n}}$.

\begin{figure*}[hbtp]
    \centering
    ~\qquad%
    \subfloat[Subcircuit $\subr{\theta_{\n}}$]{\label{fig:subr_gamma}\include{figures/Sub_routine_theta}}
    \hfill
    \subfloat[Subcircuit $\CChain{1}$]{\label{fig:Cnot chain}\include{figures/CNOT_chain_NN}}
    \hfill
    \subfloat[Subcircuit $\copygate{}{}$]{\label{fig:Cnot copy gate}\include{figures/Cnot_copysub}}
    \qquad~\\
    \subfloat[Subcircuit $\staircase{\n}$ and its recursive property]{\label{fig:staircase}\include{figures/stair_case}}
    \caption{Subcircuits for Dicke states preparation.}
    \label{fig:subroutines}
\end{figure*}

\subsection{Unbalanced Dicke states with $\excitations=1$ excitation}
\label{section:dicke_one_unbalanced}
A first generalization of $\ket{\DickeOne{\n}}$ can be obtained by removing the symmetry constraint on the coefficients in \cref{eq:dicke one def}:
\begin{equation}\label{eq:dicke one unbalanced def}
    \ket{\DickeOneU{\n}{\coeffDicke{}}} = \sum_{\elle=0}^{\n-1} \coeffDicke{\elle}\ket{2^\elle},
\end{equation}
where $\coeffDicke{\elle} \equiv \abs{\coeffDicke{\elle}} e^{i\eta_\elle}$ and $\normtwo{\coeffDicke{}}=1$.
Such a state can be prepared starting from the circuit in \cref{fig:staircase}, where the phase gates $P_{\eta_\elle}$ are defined as:
\begin{equation}\label{eq:def_phase_gate}
    \phasegate{\eta_\elle} = \begin{bmatrix}
    1 & 0 \\ 0 & e^{i \eta_\elle}
\end{bmatrix}
\end{equation}
and the angles $\theta_\elle$ are given by:
\begin{equation}
    \thetah_\elle = 2\arccos{\frac{\abs{\coeffDicke{\n-\elle-1}}}{\sqrt{1-\sum_{\jz=0}^{\n-\elle-2}\abs{\coeffDicke{\jz}}^2}}}, \label{eq_theta_unbalanced_general}
\end{equation}
for $\elle = 1,2,\ldots,\n-2$ and by:
\begin{equation}
    \thetah_{\n-1} = 2\arccos\abs{\alpha_{0}}. \label{eq_theta_unbalanced_n-1}
\end{equation}
for $\elle=\n-1$.
Note that, because of the normalization constraint, the state in \cref{eq:dicke one unbalanced def} has $\n-1$ degrees of freedom, which correspond to the number of angles $\hat\theta$.
The complex phases $\eta_{\elle}$ associated with the coefficients $\coeffDicke{\elle}$ are incorporated by appending $\n-1$ phase gates $P_\eta$.

The unbalanced Dicke state $\ket{\DickeOne{\n}(\alpha)}$ can then be prepared using the following circuit:
\begin{equation}
\label{eq:def_unbalanced_dicke1_prep}
    \begin{myqcircuit}
        \lstick{\ket{0^{\otimes\n}}} & \cgate{\DickeOne{\n}(\alpha)}{DickeOnecolor} & \rstick{\ket{\DickeOne{\n}(\alpha)}}\qw
    \end{myqcircuit}
\end{equation}
where:
\begin{equation}\label{eq_dicke_unbalanced_oracle}
\begin{myqcircuit}
& {/\strut^{\n}}\qw & \cgate{\DickeOne{\n}(\alpha)}{DickeOnecolor} & \qw \\
\end{myqcircuit}
\;=\;
\begin{myqcircuit}[0.1]
& \qw & \multicgate{3}{\staircase{\n}(\thetah)}{Deltacolor} & \gate{\phasegate{\eta_{\n-1}}} & \qw \\
& \raisebox{1ex}{$\vdots$} & & & \raisebox{1ex}{$\vdots$} \\
& \qw & \ghost{\staircase{\n}(\thetah)} & \gate{\phasegate{\eta_{1}}} & \qw \\
& \gate{\Xp} & \ghost{\staircase{\n}(\thetah)} & \gate{\phasegate{\eta_{0}}} & \qw
\end{myqcircuit}
\end{equation}
The proof of correctness of this circuit construction can be found in \cref{section:unbalanced_dicke}.

\subsection{Dicke states with neighboring $\excitations=2$ excitations}
\label{section:dicke_two_kN}
We now introduce a new and more specialized class of Dicke states with $\excitations=2$ excitations and nearest-neighbor constraints. These states will play a fundamental role in implementing operators with specific neighboring constraints within the check matrix formalism.
For simplicity, we start from the balanced case:
\begin{equation}\label{eq:def_dicke2NN_prep}
    \ket{\DickeTwoNN{\n}} = \frac{1}{\sqrt{\n-1}}\sum_{\elle=0}^{\n-2} \ket{2^\elle+2^{\elle+1}}.
\end{equation}
This state can be constructed by first preparing $\ket{\DickeOne{\n-1}}$ and then applying a \cnot{}-ladder, denoted by $\CChain{1}$ and shown in \cref{fig:Cnot chain}, so that:
\begin{equation}
\label{eq:dicke2NN_prep}
\begin{myqcircuit}
        \lstick{\ket{0^{\otimes\n}}} & \cgate{\DickeTwoNN{\n}}{DickeTwokNcolor} & \rstick{\ket{\DickeTwoNN{\n}}}\qw
    \end{myqcircuit}
\end{equation}
where:
\begin{equation}\label{eq_oracle_dicke_two_NN}
\begin{myqcircuit}
& \qw {/\strut^{\n}}& \cgate{\DickeTwoNN{\n}}{DickeTwokNcolor} &\qw \\
\end{myqcircuit}
\;=\;
\begin{myqcircuit}
  &{/\strut^{\n-2}}\qw & \multicgate{1}{\staircase{\n-1}}{Deltacolor} & \multicgate{2}{\CChain{1}}{CLkcolor} &\qw \\
 & \gate{\Xp} & \ghost{\staircase{\n-1}} & \ghost{\CChain{1}} &\qw\\
& \qw & \qw & \ghost{\CChain{1}}&\qw
\end{myqcircuit}
\end{equation}
We have that each \cnot{} gate in the ladder is activated by exactly one component of the state vector:
\begin{equation}
    \ket{2^{\elle+1}} \xrightarrow{\CChain{1}} \ket{2^{\elle} + 2^{\elle+1}},
\end{equation}
so that, by linearity, the entire superposition evolves into the correct target state:
\begin{align}
    \ket{\DickeOne{\n-1}}\ket{0} =&~\frac{1}{\sqrt{\n-1}}\sum_{\elle=0}^{\n-2} \ket{2^{\elle+1}}, \nonumber \\
    \xrightarrow{\CChain{1}}&~\frac{1}{\sqrt{\n-1}}\sum_{\elle=0}^{\n-2} \ket{2^{\elle}+2^{\elle+1}} = \ket{\DickeTwoNN{\n}}.
\end{align}

Similarly, a balanced Dicke state with $\excitations=2$ excitations and $\kN$-th nearest-neighbor constraints is defined as:
\begin{equation}
    \ket{\DickeTwokN{\n}{\kN}} = \frac{1}{\sqrt{\n-\kN}}\sum_{\elle=0}^{\n-\kN-1} \ket{2^\elle+2^{\elle+\kN}}.
\end{equation}
\Cref{eq:dicke2NN_prep} can then be generalized into the following circuit by employing the subcircuits $\staircase{\n - \kN}$ and $\CChain{\kN}$, which consists of $\n-\kN$ \cnots{} of length $\kN$:
\begin{equation}
\label{eq:dicke2kN_prep}
\begin{myqcircuit}
        \lstick{\ket{0^{\otimes\n}}} & \cgate{\DickeTwokN{\n}{\kN}}{DickeTwokNcolor} & \rstick{\ket{\DickeTwokN{\n}{\kN}}}\qw
    \end{myqcircuit}
\end{equation}
where:
\begin{equation}
\label{eq:dicke2kN_prep_circuit}
\begin{myqcircuit}
 &\qw {/\strut^{\n}}& \cgate{\DickeTwokN{\n}{\kN}}{DickeTwokNcolor} &\qw \\
\end{myqcircuit}
\;=\;
\begin{myqcircuit}
  & {/\strut^{\n-\kN-1}}\qw & \qw & \multicgate{1}{\staircase{\n-\kN}}{Deltacolor} & \multicgate{2}{\CChain{\kN}}{CLkcolor} &\qw \\
 & \gate{\Xp} & \qw & \ghost{\staircase{\n-\kN}} & \ghost{\CChain{\kN}} &\qw\\
& {/\strut^{\kN}}\qw & \qw &\qw & \ghost{\CChain{\kN}}&\qw
\end{myqcircuit}
\end{equation}

The circuit for preparing the unbalanced $\kN$-th nearest-neighboring Dicke state:
\begin{equation}
    \ket{\DickeTwokNU{\n}{\kN}{\pscoeff}} = \sum_{\elle = 0} ^{\n-\kN-1} \coeffDicke{\elle}\ket{2^{\elle}+2^{\elle+\kN}},
\end{equation}
follows immediately by substituting the subcircuit $\staircase{\n-\kN}$ with $\staircase{\n-\kN}(\thetah)$, where the angles $\thetah$ are given in \cref{eq_theta_unbalanced_general,eq_theta_unbalanced_n-1}.

\subsection{Double Dicke states}
\label{section:dicke_double}
Building upon the Dicke states defined in the previous subsections, we now introduce their \textit{double} variants by entangling them with a second register of $\n$ qubits.
We will use these states to account for the Pauli-$Y$ terms in the Hamiltonians under consideration. This is because, within the check matrix formalism, such terms correlate identical states in the $X$ and $Z$ ancillary registers, as is explained in \cref{section:main_result}.

\begin{figure*}[hbtp]
    \centering
    ~\hfill%
    \subfloat[Circuit for $\DickeTwoNN{5}$\label{fig:DickeNN for n=5}]{%
        \centering
        \include{figures/dicke5_NN}
    }%
    \hfill%
    \subfloat[Circuit for $\DickeTwokN{5}{2}$\label{fig:Dicke2N for n=5}]{%
        \centering
        \include{figures/dicke5_2N}
    }%
    \hfill%
    \subfloat[Circuit for $\DickeTwokN{5}{3}$\label{fig:Dicke3N for n=5}]{%
        \centering
        \include{figures/dicke5_3N}
    }%
    \hfill~\\
    ~\hfill%
     \subfloat[Circuit for $\DickeOneD{5}$\label{fig:DickeD for n=5}]{%
$\begin{myqcircuit}[0]
\lstick{\ket{0}} & \qw & \multicgate{4}{\staircase{5}}{Deltacolor}  & \ctrl{5} &\qw&\qw &\qw &\qw&\qw  \\
\lstick{\ket{0}} & \qw &\ghost{\staircase{5}} & \qw & \ctrl{5} & \qw &\qw &\qw &\qw \\
\lstick{\ket{0}} &\qw &\ghost{\staircase{5}} & \qw & \qw & \ctrl{5}&\qw &\qw &\qw \\
\lstick{\ket{0}} & \qw & \ghost{\staircase{5}} &\qw &\qw &\qw &\ctrl{5} &\qw &\qw \\
\lstick{\ket{0}} & \gate{\Xp} & \ghost{\staircase{5}} &\qw &\qw & \qw & \qw & \ctrl{5} &\qw \\
\lstick{\ket{0}} &\qw &\qw &\targ{} &\qw & \qw & \qw &\qw   &\qw \\
\lstick{\ket{0}} &\qw &\qw &\qw &\targ{} & \qw & \qw &\qw &\qw \\
\lstick{\ket{0}} &\qw &\qw &\qw& \qw &\targ{}& \qw &\qw &\qw \\
\lstick{\ket{0}} &\qw &\qw &\qw&\qw & \qw & \targ{} & \qw &\qw \\
\lstick{\ket{0}} &\qw &\qw &\qw&\qw & \qw & \qw & \targ{} &\qw 
\end{myqcircuit}$
     }%
     \hfill%
     \subfloat[Circuit for $\DickeTwoNND{5}$\label{fig:DickeDNN for n=5}]{%
$\begin{myqcircuit}[0]
\lstick{\ket{0}} & \qw & \multicgate{3}{\staircase{4}}{Deltacolor} &\qw &\qw &\qw &\ctrl{1} & \ctrl{5}&\qw&\qw &\qw &\qw&\qw  \\
\lstick{\ket{0}} & \qw &\ghost{\staircase{4}} & \qw &\qw&\ctrl{1}&\targ{}& \qw &\ctrl{5} & \qw &\qw &\qw &\qw \\
\lstick{\ket{0}} &\qw &\ghost{\staircase{4}} & \qw& \ctrl{1}&\targ{} &\qw &\qw & \qw & \ctrl{5}&\qw &\qw &\qw \\
\lstick{\ket{0}} & \gate{\Xp} & \ghost{\staircase{4}} & \ctrl{1} &\targ{}&\qw &\qw &\qw &\qw&\qw &\ctrl{5} &\qw &\qw \\
\lstick{\ket{0}} & \qw&\qw &\targ{} &\qw &\qw & \qw &\qw& \qw &\qw &\qw & \ctrl{5} &\qw \\
\lstick{\ket{0}} &\qw &\qw &\qw &\qw &\qw &\qw &\targ{} &\qw & \qw & \qw &\qw   &\qw \\
\lstick{\ket{0}}  &\qw &\qw &\qw &\qw &\qw &\qw &\qw &\targ{} & \qw & \qw &\qw &\qw \\
\lstick{\ket{0}}  &\qw &\qw &\qw &\qw &\qw &\qw &\qw& \qw &\targ{}& \qw &\qw &\qw \\
\lstick{\ket{0}} &\qw &\qw &\qw &\qw  &\qw &\qw &\qw&\qw & \qw & \targ{} & \qw &\qw \\
\lstick{\ket{0}} &\qw &\qw  &\qw &\qw &\qw &\qw &\qw&\qw & \qw & \qw & \targ{} &\qw 
\end{myqcircuit}$
     }%
    \hfill~
    \caption{Circuits for the preparation of Dicke states with $\n = 5$.}
    \label{fig:explicit_example}
\end{figure*}

More specifically, the balanced \emph{double Dicke} state with $\excitations=1$ excitation is defined as:
\begin{equation}
    \ket{\DickeOneD{\n}} = \frac{1}{\sqrt{\n}} \sum_{\elle=0}^{\n-1} \ket{2^\elle}\ket{2^\elle},
\end{equation}
This state can be constructed by first preparing the single-excitation balanced Dicke state $\ket{\DickeOne{\n}}$ on the first register of $\n$ qubits, and then entangling it with the second register via $\n$ \cnots{}, each acting between corresponding qubits of the two registers. Hence:
\begin{equation}\label{eq:dicke1d_prep}
    \begin{myqcircuit}
        \lstick{\ket{0^{\otimes 2\n}}} & \cgate{\DickeOneD{\n}}{DickeOneDcolor} & \rstick{\ket{\DickeOneD{\n}}}\qw
    \end{myqcircuit}
\end{equation}
where:
\begin{equation}\label{eq:dicke_one_double_oracle}
\begin{myqcircuit}
& \qw {/\strut^{2\n}\;} & \cgate{\DickeOneD{\n}}{DickeOneDcolor} &\qw \\
\end{myqcircuit}
\;=\;
\begin{myqcircuit}
&{/\strut^{\n-1}}\qw &  \multicgate{1}{\staircase{\n}}{Deltacolor} & \multicgate{2}{\copygate}{ECcolor} & \qw\\
&\gate{\Xp} &\ghost{\staircase{\n}} & \ghost{\copygate} &\qw \\
&{/\strut^{\n}}\qw & \qw & \ghost{\copygate} &\qw 
\end{myqcircuit}
\end{equation}
and where \copygate{} stands for \textit{element-wise} \cnot{} and is defined in \cref{fig:Cnot copy gate}.
Note that all $\n$ \cnots{} in the \copygate{} subcircuit can be performed in parallel.

We also define the double version of the $\ket{\DickeTwokN{\n}{\kN}}$ states. In particular, we define:
\begin{equation} \label{eq:dicke_2kN_double_prep}
    \ket{\DickeTwokND{\n}{\kN}} = \frac{1}{\sqrt{\n-\kN}} \sum_{\elle=0}^{\n-\kN-1} \ket{2^\elle+2^{\elle+\kN}}\ket{2^{\elle}+2^{\elle+\kN}},
\end{equation}
where the state $\ket{\DickeTwoNND{\n}}$ is the special case $\kN=1$.
A $\ket{\DickeTwokND{\n}{\kN}}$ state can be prepared with the same strategy as $\ket{\DickeOneD{\n}}$ by first preparing \cref{eq:dicke2kN_prep_circuit} and appending an \copygate{} entangling gate:
\begin{equation}
    \begin{myqcircuit}
        \lstick{\ket{0^{\otimes 2\n}}} & \cgate{\DickeTwokND{\n}{\kN}}{DickeTwokNDcolor} & \rstick{\ket{\DickeTwokND{\n}{\kN}}}\qw
    \end{myqcircuit}
\end{equation}
where:
\begin{equation}\label{eq:dicke_two_double_oracle}
\hspace{-0.2cm}
\begin{myqcircuit}
& \qw{/\strut^{2\n}\;} & \cgate{\DickeTwokND{\n}{\kN}}{DickeTwokNDcolor} &\qw \\
\end{myqcircuit}
\;=\;
\begin{myqcircuit}
   &{/\strut^{\n-\kN-1}\,}\qw & \qw & \multicgate{1}{\staircase{\n-\kN}}{Deltacolor} & \multicgate{2}{\CChain{\kN}}{CLkcolor} & \multicgate{3}{\copygate{}}{ECcolor} & \qw \\
&\gate{\Xp} & \qw &\ghost{\staircase{\n-\kN}} & \ghost{\CChain{\kN}} & \ghost{\copygate{}} &\qw \\
&{/\strut^{\kN}}\qw & \qw & \qw & \ghost{\CChain{\kN}} & \ghost{\copygate{}} & \qw \\
 & {/\strut^{\n}}\qw & \qw & \qw & \qw & \ghost{\copygate{}} &\qw 
\end{myqcircuit}
\end{equation}
For clarity, we present an example in \cref{fig:explicit_example} that illustrates the circuit structure for the simple case of $\n = 5$.

The circuits for preparing the unbalanced double Dicke states:
\begin{align}
    \ket{\DickeOneUD{\n}{\pscoeff}} &= \sum_{\elle = 0} ^{\n-1} \coeffDicke{\elle}\ket{2^{\elle}}\ket{2^{\elle}}, \\
    \ket{\DickeTwokNUD{\n}{\kN}{\pscoeff}} &= \sum_{\elle = 0} ^{\n-\kN-1} \coeffDicke{\elle}\ket{2^{\elle}+2^{\elle+\kN}}\ket{2^{\elle}+2^{\elle+\kN}},
\end{align}
follow again from substituting the subcircuits $\staircase{\n}$ and $\staircase{\n-\kN}$ with $\staircase{\n}(\thetah)$ and $\staircase{\n-\kN}(\thetah)$, respectively, where the angles $\thetah$ are given in \cref{eq_theta_unbalanced_general,eq_theta_unbalanced_n-1}.

\section{Applications to spin models}
\label{section:applications}

The main contribution of this paper was presented in \cref{section:main_result}: a significantly simpler and more compact implementation of the $\select$ oracle in the \cmlcu{} framework. This improvement is made possible by the use of the check matrix formalism and the introduction of additional ancilla qubits (specifically, $2\n$).
At first glance, this might suggest an exponential cost for the $\PR$ and $\PL$ oracles, given that generic state preparation scales exponentially with the number of qubits \cite{Plesch2011}.

In this section, however, we demonstrate that for certain models, the total \cmlcu{} cost can be reduced to linear in the number of LCU terms by exploiting the sparsity and structure of the Hamiltonian coefficients $\newpscoeff$. To clearly illustrate our approach, we focus on two 1D spin systems with open boundary conditions: first, the Heisenberg model in \cref{section:heisenberg}, followed by the spin glass model in \cref{section:spin_glass}.
While our approach centers on these 1D examples, the techniques and results are broadly applicable to higher dimensions, as discussed in \cref{section:generalization,app: generalization}.

\subsection{Heisenberg model}
\label{section:heisenberg}
We consider the Heisenberg model for quantum spins on a one-dimensional chain. This prototypical model captures the effects of strong interactions between electronic spins and is commonly used in condensed matter physics for benchmarking new computational techniques. The one-dimensional Heisenberg Hamiltonian with open boundary conditions is defined as:
\begin{equation}\label{eq:def_ham_heisenberg}
\begin{split}
\Ham =& \sum_{\elle=0}^{\n-1} \gx \Xp_\elle + \gz \Zp_\elle + \gy \Yp_\elle \\
 +& \sum_{\elle=0}^{\n-2} \Jx \Xp_\elle\Xp_{\elle+1} + \Jz\Zp_\elle\Zp_{\elle+1} + \Jy \Yp_\elle\Yp_{\elle+1},
\end{split}
\end{equation}
where $g^\zeta$ and $J^\zeta$, for $\zeta \in \{x,y,z\}$, denote the external magnetic field and the interaction between neighboring spins, respectively.
Notably, this Hamiltonian is already expressed as a linear combination of Pauli strings, as in \cref{eq:def_lcu_pauli}.

To construct an efficient \cmlcu{} block encoding for the Heisenberg Hamiltonian given in \cref{eq:def_ham_heisenberg}, we start by mapping each term to a coefficient-state pair of the form $(\newpscoeff_{\ix\jz}, \, \ket{\ix}\ket{\jz})$.
Recall that the first register $\ket{\ix}$ controls the activation of the $\Xp$ contribution during the $\select$ procedure, while the second register $\ket{\jz}$ activates the application of $\Zp$.
For example, the term $\gx\Xp_{\elle}$ can naturally be mapped to $(\gx,\,\ket{2^{\elle}}\ket{0})$:
\begin{align}
    \label{eq:example_state_matrix_x}
    \gx\Xp_{\elle} &= \underbrace{\Ip\otimes\cdots\otimes\Ip}_{\n-1-\elle}\otimes \left(\gx\Xp \right)\otimes\underbrace{\Ip\otimes\cdots\otimes\Ip}_{\elle} \nonumber \\
    &\rightarrow (\gx,\, \ket{\underbrace{0\cdots0}_{\n-1-\elle} 1 \underbrace{0\cdots0}_{\elle}} \otimes \ket{0^{\otimes \n}}) \nonumber \\
    &= (\gx,\, \ket{2^{\elle}}\ket{0}).
\end{align}
Analogously, the term $\gz\Zp_{\elle}$ is mapped to $(\gz,\, \ket{0}\ket{2^\elle})$.
In contrast, the term $\gy\Yp_{\elle}$ always contributes equally to both $\Xp$ and $\Zp$, and is therefore associated with the mapping:
\begin{align}
    \label{eq:example_state_matrix_y}
    \gy\Yp_{\elle} &= \underbrace{\Ip\otimes\cdots\otimes\Ip}_{\n-1-\elle}\otimes \left(-i\gy\Zp_\elle\Xp_\elle \right)\otimes\underbrace{\Ip\otimes\cdots\otimes\Ip}_{\elle} \nonumber \\
    &\rightarrow (-i\gy,\, \ket{\underbrace{0\cdots0}_{\n-1-\elle}1\underbrace{0\cdots0}_{\elle}\ }\otimes \ket{\underbrace{0\cdots0}_{\n-1-\elle}1\underbrace{0\cdots0}_{\elle}}) \nonumber \\
    &= (-i\gy,\, \ket{2^{\elle}}\ket{2^{\elle}}).
\end{align}
The nearest-neighbor terms can be constructed following the same reasoning, and \cref{tab:ancilla_operators} summarizes the mapping for all terms in \cref{eq:def_ham_heisenberg}.

\begin{table}
\centering
\begin{tabular}{|c|c|c|}
\hline
\textbf{Operator} & $\boldsymbol{\newpscoeff_{\ix\jz}}$ & $\mathbf{\ket{\ix}\ket{\jz}}$  \\
\hline
$\gx\Xp_\elle$ & $\gx$ & $\ket{2^{\elle}}\ket{0}$ \\
\hline
$\gz\Zp_\elle$ & $\gz$ & $\ket{0}\ket{2^{\elle}}$ \\
\hline
$\gy\Yp_\elle$ & $-i\gy$ & $\ket{2^{\elle}}\ket{2^{\elle}}$ \\
\hline
$\Jx\Xp_\elle\Xp_{\elle+1}$ & $\Jx$ & $\ket{2^\elle+2^{\elle+1}}\ket{0}$ \\
\hline
$\Jz\Zp_\elle\Zp_{\elle+1}$ & $\Jz$ & $\ket{0}\ket{2^\elle+2^{\elle+1}}$ \\
\hline
$\Jy\Yp_\elle\Yp_{\elle+1}$ & $-\Jy$ & $\ket{2^\elle+2^{\elle+1}}\ket{2^\elle+2^{\elle+1}}$ \\
\hline
\end{tabular}
\caption{The mapping of each Pauli string in the Heisenberg Hamiltonian given in \cref{eq:def_ham_heisenberg} to the corresponding coefficient and ancilla state pair as defined in \cref{eq:lcu_newpscoeff,eq:lcu_general_check_matrix}.}
\label{tab:ancilla_operators}
\end{table}

Once each Pauli string of the Heisenberg Hamiltonian is mapped to the corresponding state following the check matrix formalism of \cref{eq:lcu_newpscoeff,eq:lcu_general_check_matrix}, we focus on implementing the state preparation oracles in \cref{fig:be_general}. Following \cref{eq:PR_state_general}, the $\PR$ oracle must prepare the following complex state on the $2\n$ ancillae:
\begin{align}
\frac{1}{\sqrt{\normalizationDicke}}\Bigg[\sum_{\elle=0}^{\n-1} &\sqrt{\gx}\ket{2^\elle}\ket{0} + \sqrt{\gz}\ket{0}\ket{2^\elle} + \sqrt{-i\gy}\ket{2^\elle}\ket{2^\elle} \nonumber \\
+\sum_{\elle=0}^{\n-2} &\Big(\sqrt{\Jx}\ket{2^\elle+2^{\elle+1}}\ket{0} + \sqrt{\Jz}\ket{0}\ket{2^\elle+2^{\elle+1}} \nonumber \\
&+\sqrt{-\Jy}\ket{2^\elle+2^{\elle+1}}\ket{2^\elle+2^{\elle+1}}\!\Big)\Bigg],
\label{eq:prep_state_heisenberg}
\end{align}
where $\normalizationDicke$ is the same normalization factor as for the standard LCU:
\begin{equation}
\begin{split}
    \normalizationDicke &= \n\left(\abs{\gx}+\abs{\gy}+\abs{\gz}\right) \\
    &+(\n-1)\left(\abs{\Jx}+\abs{\Jy}+\abs{\Jz}\right).
\end{split}
\end{equation}
Similarly, the $\PL$ oracle prepares the same state but with complex conjugate coefficients.

Remark that the state in \cref{eq:prep_state_heisenberg} is very sparse: only $6\n-3$ entries among the $4^n$ ones of a $2\n$-qubits state are nonzero.
By exploiting this sparsity structure, we will now introduce a new and efficient circuit for the Heisenberg $\PR$ and $\PL$ oracles. The implementation is based on Dicke states and scales as $\bigO{\n}$. This is in contrast to using existing sparse state preparation algorithms~\cite{Malvetti2021, Zhang2022,Li2025}, that typically scale from $\bigO{\sparseindex\n}$ to $\bigO{\sparseindex\n/\log{n}}$ if we allow for extra ancillaries, where $\sparseindex$ is the sparsity.

By grouping together the one-body and two-body terms with the same coefficients, and using the definitions of balanced Dicke and double Dicke states, we can reformulate \cref{eq:prep_state_heisenberg} as follows:
\begin{align}
    \frac{1}{\sqrt{\normalizationDicke}} \Big(&\sqrt{\n\gx}\ket{\DickeOne{\n}}\ket{0} + \sqrt{\n\gz}\ket{0}\ket{\DickeOne{\n}} + \sqrt{-i\n\gy}\ket{\DickeOneD{\n}} \nonumber \\
    +\,&\sqrt{(\n-1)\Jx}\ket{\DickeTwoNN{\n}}\ket{0} + \sqrt{(\n-1)\Jz}\ket{0}\ket{\DickeTwoNN{\n}} \nonumber \\[0.5ex]
    +\,&\sqrt{-(\n-1)\Jy}\ket{\DickeTwoNND{\n}}\!\Big).
\label{eq:prep_state_heisenberg_dicke}
\end{align}
Quantum circuits for the Dicke states $\ket{\DickeOne{\n}}$, $\ket{\DickeOneD{\n}}$, $\ket{\DickeTwoNN{\n}}$, and $\ket{\DickeTwoNND{\n}}$ are given in \cref{section:dicke}.

\begin{figure}[hbtp]
    \centering
    \hspace*{2.25em}\input{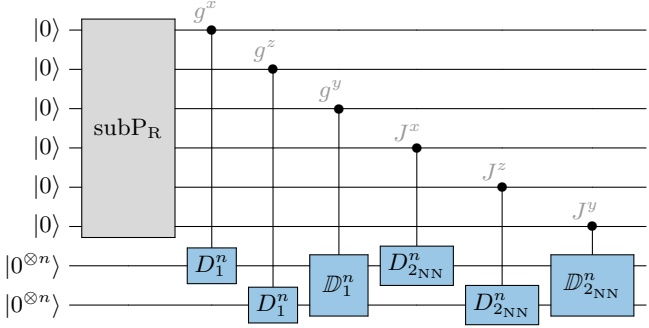}\\
    \caption{High-level circuit implementation of the $\PR$ oracle for the Heisenberg model. The $\PL$ oracle can be constructed in an analogous way by considering the complex conjugate of the coefficients in the $\subPR$.}
    \label{fig:heisenberg_prep}
\end{figure}

\begin{figure}[htbp]
    \centering
    \input{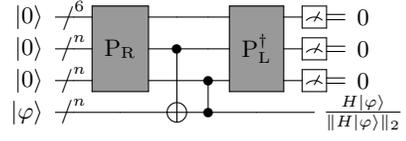}
    \caption{\cmlcu{} circuit for the Heisenberg model. The \PR{} and \PL{} oracles act on $2\n + 6$ ancillae, while the \select{} consists of $\n$ \cnot{} and
$\n$ \cz{} gates, as in \cref{fig:be_general}.}
    \label{fig:be_heisenberg}
\end{figure}

Then, it remains to efficiently implement the 6 scalar values in \cref{eq:prep_state_heisenberg_dicke} containing the Heisenberg parameters. Although 3 extra ancillae would be enough for encoding 6 values, we opt to employ 1 ancilla per value to allow for single-controlled Dicke circuits, as visualized in \cref{fig:heisenberg_prep}.
Consequently, this brings the total number of qubits for the \cmlcu{} block encoding of the Heisenberg model to $3n+6$, as depicted in \cref{fig:be_heisenberg}.

The subcircuit $\subPR$ prepares then the following unbalanced Dicke state with $\excitations=1$:
\begin{equation}
\label{eq:def_subPR}
\begin{split}
   \sqrt{\nicefrac{  \n\gx}{\normalizationDicke}}    & \ket{100000} \\
+\,\sqrt{\nicefrac{  \n\gz}{\normalizationDicke}}    & \ket{010000} \\
+\,\sqrt{\nicefrac{-i\n\gy}{\normalizationDicke}}    & \ket{001000} \\
+\,\sqrt{\nicefrac{ (\n-1)\Jx}{\normalizationDicke}} & \ket{000100} \\
+\,\sqrt{\nicefrac{ (\n-1)\Jz}{\normalizationDicke}} & \ket{000010} \\
+\,\sqrt{\nicefrac{-(\n-1)\Jy}{\normalizationDicke}} & \ket{000001},
\end{split}
\end{equation}
which can be implemented by the circuit given in \cref{eq_dicke_unbalanced_oracle}.

\begin{figure*}[hbtp!]
    \centering
    \hspace*{3.5em}\input{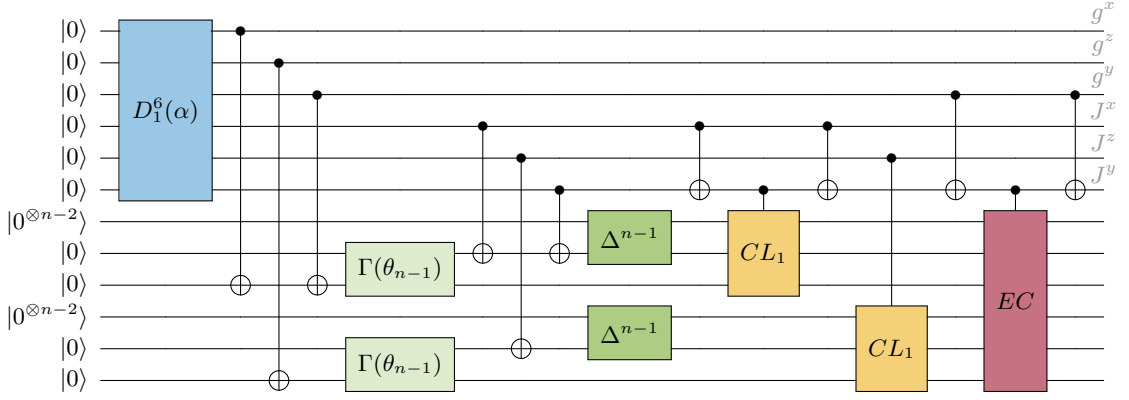}
    \caption{Compact implementation of the $\PR$ oracle for the Heisenberg model with $\alpha$ given in \cref{eq:def_subPR} and $\theta_{\n-1}$ in \cref{eq:def_theta_dicke1}.}
    \label{fig:compact_heisenberg}
\end{figure*}

Although the circuit in \cref{fig:heisenberg_prep} already provides a relatively efficient state preparation strategy, the specific structure of the Heisenberg model allows for further simplification, as elaborated in \cref{subsec: compression Heisenberg}.
This optimization leverages both the structure of the state prepared by \subPR{} and the properties of the Dicke state circuits discussed in \cref{section:dicke}.
The resulting optimized circuit is shown in \cref{fig:compact_heisenberg}.

Finally, the $\PL$ oracle is defined analogously with its subroutine $\subPL$ preparing the conjugate of \cref{eq:def_subPR}.

~

~

\subsection{Spin glass model}
\label{section:spin_glass}
We now generalize the one-dimensional Heisenberg model to the case of non-uniform, position-dependent spin couplings and transverse fields, introducing all-to-all interactions. Such a system is often referred to as a \textit{spin glass} model \cite{Edwards_1975,Kirkpatrick1978,Binder1986}, characterized by its rugged energy landscape that makes finding low-energy states computationally challenging. Recent advancements suggest that quantum algorithms can efficiently explore local minima on these complex landscapes, offering a potential quantum advantage in solving such problems \cite{Chen2025}. 
To the best of our knowledge, \cite{sanders2020} presents the only prior work on the block encoding of spin-glass-like Hamiltonians for combinatorial optimization, with a particular focus on the commuting case.

We consider the most general definition of a spin glass Hamiltonian:
\begin{widetext}
\begin{equation}
\label{eq:def_ham_spin_glass}
   \Ham = \sum_{\elle=0}^{\n-1} \gx_{\elle}\Xp_{\elle} + \gz_{\elle}\Zp_{\elle} + \gy_{\elle}\Yp_{\elle}
   +\sum_{\kN=1}^{\n-1}\sum_{\elle=0}^{\n-\kN-1}\Jx_{\elle,\elle+\kN}\Xp_{\elle}\Xp_{\elle+\kN} + \Jz_{\elle,\elle+\kN}\Zp_{\elle}\Zp_{\elle+\kN} + \Jy_{\elle,\elle+\kN}\Yp_{\elle}\Yp_{\elle+\kN},
\end{equation}
\end{widetext}
where $\kN\in\{1,\ldots,\n-1\}$ denotes the separation between the operators, i.e.:
\begin{equation*}
    \Xp_{\elle}\Xp_{\elle+\kN} = \underbrace{\Ip \otimes \cdots \otimes \Ip}_{\n-1-\elle-\kN}\otimes \Xp\otimes \underbrace{\Ip \otimes \cdots \otimes \Ip}_{\kN-1}\otimes \Xp \otimes\underbrace{\Ip \otimes \cdots \otimes \Ip}_{\elle},
\end{equation*}
so that $\kN=1$ corresponds to the nearest-neighbor case.

To block encode such a Hamiltonian using the \cmlcu{} method, we construct the \PR{} and \PL{} subcircuits in a manner similar to those described in \cref{section:heisenberg}, while the \select{} subcircuit remains unchanged by design. Therefore, we now generalize the circuit shown in \cref{fig:heisenberg_prep} to accommodate long-range and non-uniform interactions, while continuing to leverage the structure of Dicke states.

We begin by applying the same operator-state mapping logic outlined in \cref{section:heisenberg}, with a particular focus on the mapping detailed in \cref{tab:ancilla_operators}.
Consequently, the $\PR$ oracle must prepare the following state:
\begin{widetext}
\begin{equation}
\label{eq:Ham_state}
\begin{split}
   \frac{1}{\sqrt{\generalnorm}} \Bigg[& \sum_{\elle=0}^{\n-1} \sqrt{\gx_{\elle}}\ket{2^{\elle}}\ket{0} +\sqrt{\gz_{\elle}}\ket{0}\ket{2^{\elle}} +\sqrt{-i\gy_{\elle}}\ket{2^{\elle}}\ket{2^{\elle}} \\
   +&\sum_{\kN=1}^{\n-1}\sum_{\elle=0}^{\n-\kN-1} \sqrt{\Jx_{\elle,\elle+\kN}}\ket{2^{\elle}+2^{\elle+\kN}}\ket{0} + \sqrt{\Jz_{\elle,\elle+\kN}}\ket{0}\ket{2^{\elle}+2^{\elle+\kN}} + \sqrt{-\Jy_{\elle,\elle+\kN}}\ket{2^{\elle}+2^{\elle+\kN}}\ket{2^{\elle}+2^{\elle+\kN}} \Bigg] ,
\end{split}
\end{equation}
\end{widetext}
where $N$ is again the same normalization factor as for the standard LCU:
\begin{equation}
\begin{split}
    \generalnorm &= \sum_{\elle=0}^{\n-1} \left( \vert \gx_{\elle} \vert + \vert \gz_{\elle} \vert + \vert \gy_{\elle} \vert\right) \\
    &+ \sum_{\kN=1}^{\n-1}\sum_{\elle=0}^{\n-\kN-1} \left(\vert \Jx_{\elle,\elle+\kN} \vert + \vert \Jz_{\elle,\elle+\kN} \vert + \vert \Jy_{\elle,\elle+\kN} \vert \right).
\end{split}
\end{equation}

Note that the state in \cref{eq:Ham_state} shares a similar structure with \cref{eq:prep_state_heisenberg}, but includes longer-range interaction terms and non-uniform coefficients. 
As a consequence, we need to substitute the balanced Dicke states from \cref{eq:prep_state_heisenberg_dicke} with their unbalanced and long range counterparts, $\ket{\DickeOne{\n}(\alpha)}$, $\ket{\DickeOneD{\n}(\alpha)}$, $\ket{\DickeTwokN{\n}{\kN}(\alpha)}$, and $\ket{\DickeTwokND{\n}{\kN}(\alpha)}$.

In order to do so, we first define a coefficient matrix for each dimension $\zeta \in \{\xsub,\ysub,\zsub\}$.
In particular, for $\xsub$ we define the entries:
\begin{equation}
\label{eq:coefmatrix_cases}
\coefmat{\xsub}_{(\elle,\jz)}= \begin{cases}
0 & \text{if $\elle>\jz$} \\
\sqrt{\gx_{\elle}} & \text{if $\elle=\jz$}\\
\sqrt{\Jx_{\elle,\jz}} & \text{if $\elle <\jz$}
\end{cases}.
\end{equation}
so that the matrix $\coefmat{\xsub}$ takes an upper diagonal form:
\begin{equation}
\label{eq:coefmatrix explicit}
    \coefmat{\xsub} = \bigcoefmat{\xsub}.
\end{equation}
The matrices $\coefmat{\zsub}$ and $\coefmat{\ysub}$ are defined analogously. 
This notation is particularly convenient because the terms in \cref{eq:Ham_state} involving $\kN$-th nearest-neighbor interactions are directly mapped to the matrix elements on the $\kN$-th upper diagonal of $\coefmat{\xsub}$, which we denote as $\coefmatk{\xsub}$.
Extending $[\kN]$ to include $\kN=0$ corresponds to incorporating the main diagonal, which encodes the one-body terms.

Since each of the corresponding unbalanced Dicke states must be normalized, the same requirement applies to the elements on every $\kN$-th upper diagonal.
For example, we define the normalization factor for $\xsub$ with $\kN\in\{0,\ldots,\n-1\}$ as:
\begin{equation}
    \label{eq:norm_diagonal}
    \diagnorm{\xsub}{\kN} = \sum_{\elle=0}^{\n-\kN-1} \vert\coefmat{\xsub}_{(\elle,\elle+\kN)}\vert^2,
\end{equation}
yielding the normalized coefficient vectors:
\begin{equation}
    \coefmatnormk{\xsub} = \frac{\coefmatk{\xsub}}{\sqrt{\diagnorm{\xsub}{\kN}}}.
\end{equation}
The state in \cref{eq:Ham_state} can then be rewritten as:
\begin{widetext}
\begin{equation}\label{eq:subpr_spin_glass}
\begin{split}
    \frac{1}{\sqrt{\generalnorm}} \Bigg[&\sqrt{\diagnorm{\xsub}{0}}\ket{\DickeOneU{\n}{\coefmatnormzero{\xsub}}}\ket{0} 
    + \sqrt{\diagnorm{\zsub}{0}}\ket{0}\ket{\DickeOneU{\n}{\coefmatnormzero{\zsub}}} + \sqrt{-i\diagnorm{\ysub}{0}} \ket{\DickeOneUD{\n}{\coefmatnormzero{\ysub}}} \\
    +&\sum_{\kN=1}^{\n-1} \sqrt{\diagnorm{\xsub}{\kN}} \ket{\DickeTwokNU{\n}{\kN}{\coefmatnormk{\xsub}}}\ket{0} + \sqrt{\diagnorm{\zsub}{\kN}}\ket{0}\ket{\DickeTwokNU{\n}{\kN}{\coefmatnormk{\zsub}}} 
    + \sqrt{-\diagnorm{\ysub}{\kN}}\ket{\DickeTwokNUD{\n}{\kN}{\coefmatnormk{\ysub}}} \Bigg].
\end{split}
\end{equation}
\end{widetext}

Note that \cref{eq:subpr_spin_glass} resembles \cref{eq:prep_state_heisenberg_dicke}, although it contains $3\n$ terms instead of $6$.
By adopting the same strategy used for the Heisenberg model, we can define a subcircuit $\subPR$ on $3\n$ extra ancillae which prepares the following state: 
\begin{align}\label{eq:oracle_diag}
    \frac{1}{\sqrt{\generalnorm}} \Bigg[ &\sum_{\kN = 0}^{\n-1} \sqrt{\diagnorm{\xsub}{\kN}}\ket{2^{3(\n-\kN)-1}} 
    +\sum_{\kN = 0}^{\n-1} \sqrt{\diagnorm{\zsub}{\kN}}\ket{2^{3(\n-\kN)-2}} \nonumber \\
    +& \sqrt{-i\diagnorm{\ysub}{0}}\ket{2^{3\n-3}} + \sum_{\kN=1}^{\n-1}\sqrt{-\diagnorm{\ysub}{\kN}}\ket{2^{3(\n-\kN)-3}} \Bigg],
\end{align}
which, once again, corresponds to an unbalanced Dicke state with $\excitations=1$ excitation and can thereby be prepared with linear gate count.
The $\Yp$ term for $\kN = 0$ is isolated from the rest, as it carries an additional factor of $\sqrt{i}$.

\begin{figure*}[hbtp]
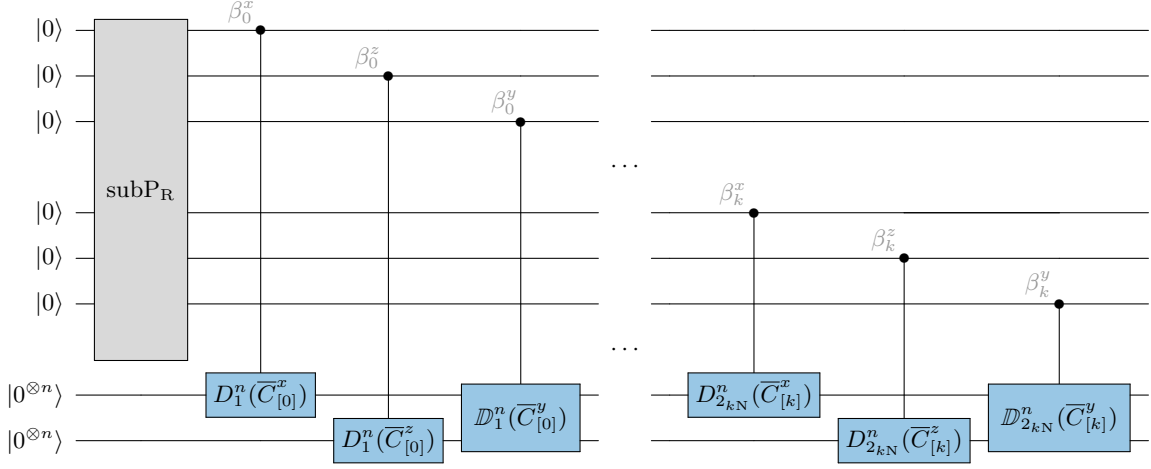

    \centering
    \include{figures/spin_glass_prep}
    \caption{Circuit implementing $\PR$ for the spin glass model. $\PL$ can be constructed in an analogous way by considering the complex conjugate of the coefficients in the $\subPR$.}
    \label{fig:spin_glass_prep}
\end{figure*}

The generalization of \cref{fig:heisenberg_prep} is shown in \cref{fig:spin_glass_prep}, which presents the circuit implementing the $\PR$ oracle for the spin glass Hamiltonian.
As mentioned earlier, the $\PL$ oracle prepares the complex conjugate of the state in \cref{eq:subpr_spin_glass} and has the same circuit structure as $\PR$.
It is important to note that, for a structureless Hamiltonian, the Dicke state subcircuits in \cref{fig:spin_glass_prep} cannot be grouped together or simplified as in \cref{fig:compact_heisenberg}, due to the varying rotation angles $\theta$ across different blocks.
In this case, the spin glass \cmlcu{} block encoding requires $5\n$ ancillae and $\bigO{\n^2}$ gates.

On the other hand, if the system exhibits additional structure, we can exploit it to reduce resource requirements.
For example, if the interactions decay and $\kN$ can be truncated beyond a certain cutoff $\esse$, the number of controlled Dicke state subcircuits drops to $\esse$, lowering the overall gate count to $\bigO{\n\esse}$. Similarly, the number of ancillae required for implementing $\subPR$ decreases from $3\n$ down to $3(\esse+1)$.
Moreover, if the interactions are identical across all site pairs at a fixed $\kN$, the subcircuits in \cref{fig:spin_glass_prep} can be simplified to prepare balanced Dicke states. In this case, we can compress the Dicke states subcircuits using the approach outlined in \cref{subsec: compression Heisenberg} for the Heisenberg model.
Nevertheless, even when the coefficients in \cref{eq:Ham_state} are non-constant, it remains possible to construct a more efficient implementation of \PR{} and \PL{} compared to \cref{fig:spin_glass_prep}. This improved construction is described in detail in \cref{subsec: compression Spin Glass}.

\subsection{Generalization}
\label{section:generalization}
We close \cref{section:applications} by discussing the applicability of the \cmlcu{} framework to a much broader class of spin Hamiltonians beyond the illustrative Heisenberg and spin glass models. First, we show that axis-aligned spin-spin interactions, i.e., $X_{\ell}X_{\ell
+\kN}$, $Y_{\ell}Y_{\ell + \kN}$, and $Z_{\ell}Z_{\ell + \kN}$ explored in \cref{section:heisenberg,section:spin_glass}, can be naturally generalized to arbitrary two-body interactions. In \cref{app: generalization}, we give explicit constructions of the $\PR$ oracle for representative terms, including $Y_{\ell}X_{\ell+\kN}$, $Y_{\ell}X_{\ell+\kN}$, and $X_{\ell} Z_{\ell+\kN}$, thus enabling block encoding of two-body Pauli strings with any interaction types and strengths. 

We also remark that the two-body interactions provide a basis for recursively handling three-body interactions. However, to fully preserve the favorable resource overhead of the spin glass circuit (or to remain asymptotically comparable), we require additional structure: either the interaction terms must be geometrically local (e.g. $X_{\ell} X_{\ell +\kN} X_{\ell +\kN + \kN'}$ with $\kN + \kN' = \bigO 1$), or the interaction strengths must be homogeneous within each interaction type. Similar considerations are expected to hold when extending our framework to higher-order spin interactions.

Second, while our demonstration primarily focuses on spin-1/2 Hamiltonians, the algorithm formally extends to higher-spin models in a qudit-based setting. Recall that the check matrix formalism introduced in \cref{section:main_result} maps each $n$-qubit Pauli string to a $2n$-binary pair $(i,j)$, where the $n$-bit strings $i$ and $j$ respectively indicate the presence of $X$ and $Z$ components on each qubit. For higher-spin systems, the corresponding $d$-level $X$ and $Z$ matrices \cite{Patera_1988,Nielsen_2002} can be defined as:
\begin{align}
    X_{\ell} \ket{\isbit{\ell}} &= \ket{\isbit{\ell} \oplus 1}, 
    \\
    Z_{\ell} \ket{\isbit{\ell}} &= \omega^{ \isbit{\ell}} \ket{\isbit{\ell}}, & \omega &\equiv e^{2\pi i/d},
\end{align}
where the summation symbol $\oplus$ is intended modulo $d$.
Such operators allow any $n$-qudit Pauli string to be uniquely represented as a $d$-nary (binary for $d=2$, ternary for $d=3$, etc.)~pair $(i,j)$ specifying the integer powers of $X$ and $Z$ on each site. Such a $d$-nary representation, $\bigotimes_{\elle=0}^{\n-1} \Zp^{\jzbit{\elle}}\Xp^{\ixbit{\elle}} $, can be encoded using $2n$ ancillary qudits along with elementary controlled $X$ and $Z$ operations:
\begin{align}
    \cnot \ket{\ixbit{\ell}, \isbit{\ell} } &= \ket{ \ixbit{\ell}, \ixbit{\ell} \oplus \isbit{\ell}} = X_{\ell}^{\ixbit{\ell}} \ket{\ixbit{\ell}, \isbit{\ell} }, \\
    \cz \ket{\jzbit{\ell}, \isbit{\ell}} &= \omega^{\jzbit{\ell} \isbit{\ell}} \ket{\jzbit{\ell}, \isbit{\ell}} = Z_{\ell}^{\jzbit{\ell}} \ket{\jzbit{\ell}, \isbit{\ell}},
\end{align}
which constitute our $\select$ oracle. The $(\PR, \PL)$ oracles can now be realized via the preparation of qudit Dicke states \cite{Nepomechie_2024, kerzner2025}, although their optimal implementation remains an open question.
The qudit version of our algorithm can be useful for block encoding paradigmatic higher-spin Hamiltonians, such as the spin-1 Heisenberg or Affleck-Kennedy-Lieb-Tasaki (AKLT) models~\cite{tasaki2020}.

Finally, we observe that our framework can be applied to spin systems with pairwise interactions in higher dimensions.
This is because the spins can always be relabeled in a lexicographic order, effectively mapping the system to a one-dimensional chain as in the spin glass model.
Using this mapping, it is then possible to construct block encodings for geometrically local Hamiltonians defined on higher-dimensional lattices. Although the relabeling does not preserve the geometric locality, since neighboring sites in the original lattice may map to distant ones in the 1D ordering, the total number of terms still scales linearly with system size. This sparsity ensures that block encodings of such Hamiltonians remain efficient, even in higher dimensions.

\section{End-to-end resource estimation}
\label{section:resource estimation}

We now estimate the total quantum resources required to implement \cmlcu{} for both the Heisenberg and spin glass models.
We start in \cref{section:dicke_cost} by reviewing computational cost for preparing Dicke states, which are used as subcircuits for the \cmlcu{} block encodings.
We then investigate the two spin models in \cref{section:heisenberg_cost} and \cref{section:spin_glass_cost}.
Finally, in \cref{section:comparison}, we benchmark the performance of \cmlcu{} and compare it to other state-of-the-art block-encoding methods.

As a metric for resource estimation, we consider the number of \cnots{} required in our circuits. In particular, we first count each Toffoli and two-qubit gate separately.
Then, we consider their decompositions into \cnots{} at the final stage.

\subsection{Dicke states preparation}
\label{section:dicke_cost}
In this subsection, we describe the computational cost in number of \cnot{} gates for the Dicke states preparation circuits introduced in \cref{section:dicke}. The results are summarized in \cref{tab:dicke-cost}.

We begin with the circuit for preparing the balanced Dicke state $\ket{\DickeOne{n}}$, defined in \cref{eq_oracle_dicke_one}. The most elementary subcircuit, $\subr{\theta}$, shown in \cref{fig:subr_gamma}, can be implemented with only $2$ \cnot{} gates, as proven in \cref{section: proof of 2 cnot}. Since the $\staircase{\n}$ circuit, shown in \cref{fig:staircase}, is composed of $\n - 1$ $\subr{\theta}$ subcircuits, both $\staircase{\n}$ and the preparation of $\ket{\DickeOne{\n}}$ require a total of $2\n - 2$ \cnot{} gates.
Note that preparing the unbalanced Dicke state $\ket{\DickeOne{n}(\alpha)}$, defined in \cref{eq_dicke_unbalanced_oracle}, requires only additional phase gates. Therefore, its \cnot{} cost remains the same as that of its balanced counterpart.

The preparation circuit of the nearest-neighbor Dicke state $\ket{\DickeTwoNN{\n}}$, defined in \cref{eq_oracle_dicke_two_NN}, involves a $\staircase{\n-1}$ subcircuit and an additional $\n - 1$ \cnot{} gates from the $\CChain{1}$ ladder shown in \cref{fig:Cnot chain}. This leads to a total \cnot{} cost of $3\n - 5$. Likewise, preparing the $k$th nearest-neighbor Dicke state $\ket{\DickeTwokN{\n}{\kN}}$, defined in \cref{eq:dicke2kN_prep_circuit}, requires a $\staircase{\n-\kN}$ subcircuit and $\n - \kN$ extra \cnot{} gates, yielding a total \cnot{} cost of $3\n - 3\kN - 2$.
Again, preparing the unbalanced versions does not require extra \cnot{} gates.

To prepare the double Dicke states, $\DickeOneD{\n}$ and $\DickeTwokND{\n}{\kN}$, defined in \cref{eq:dicke_one_double_oracle,eq:dicke_two_double_oracle}, respectively, it is sufficient to append an \copygate{} subcircuit, shown in \cref{fig:Cnot copy gate}, leading to $\n$ additional \cnot{} gates.

Although we focus on the \cnot{} gate count as the primary metric of computational cost, we remark that Dicke states can also be prepared with logarithmic \cite{Bartschi_2022} or even constant circuit depth \cite{Piroli2024,Buhrman2024,Farrell2025} when mid-circuit measurements are available.
While the \cnots{} in the \copygate{} subcircuit can naturally be performed in parallel, the $\CChain{\kN}$ subcircuit can also be implemented in constant depth using mid-circuit measurements \cite{Bumer2024}.
This implies that, in theory, all the aforementioned subcircuits can be realized with constant-depth circuits.

Nevertheless, we rely on the original linear-depth construction, as it is more amenable to generalization within our framework.
In particular, certain applications in \cref{section:applications} require controlled versions of these subcircuits,
which are significantly easier to implement in their linear-depth form.
We leave open the question of whether controlled Dicke state preparation circuits can also be implemented in constant depth by leveraging mid-circuit measurements.

\begin{table}
\centering
\subfloat[]{\label{tab:cnot sub}
\begin{tabular}{|c|c|}
\hline
     \textbf{Subcircuit}    &\textbf{\cnots{}}\\
     \hline
        $\subr{\theta}$& 2\\
        \hline
        $\staircase{\n}$ & $2\n-2$ \\
        \hline
        $\CChain{\kN}$ & $\n-\kN$ \\
        \hline
        $\copygate{}$ & $\n$ \\
        \hline
        \end{tabular}
}\\
    \subfloat[]{\label{tab: cNot dicke}
\begin{tabular}{|l|c|l|c|}
\hline
\multicolumn{2}{|c|}{\textbf{Dicke}} & \multicolumn{2}{c|}{\textbf{Double Dicke}} \\
\hline
\textbf{State} & \textbf{\cnots{}}  &\textbf{State} &\textbf{\cnots{}}\\
\hline
$\DickeOne{\n}$ & $2\n -2$ &$\DickeOneD{\n}$ & $3\n -2$\\
\hline
$\DickeTwoNN{\n}$ & $3\n -5$ &$\DickeTwoNND{\n}$ & $4\n -5$\\
\hline
$\DickeTwokN{\n}{\kN}$ & $3\n-3\kN-2$ &$\DickeTwokND{\n}{\kN}$& $4\n - 3\kN -2$\\
\hline
\end{tabular}}
\caption{Total \cnot{} gate counts for the subcircuits given in \cref{fig:subroutines} and the preparation circuits for Dicke states defined in \cref{section:dicke}. The preparation of unbalanced Dicke states yields the same \cnot{} gate counts as the balanced ones.\label{tab:dicke-cost}}
\end{table}

\subsection{Heisenberg \cmlcu{} block encoding}
\label{section:heisenberg_cost}
For the Heisenberg model, we analyze the \cmlcu{} circuit in \cref{fig:be_heisenberg}, where the state preparation oracles $\PR$ and $\PL$ are compactly implemented as in \cref{fig:compact_heisenberg}.
The total cost of the Heisenberg \cmlcu{} block encoding is summarized in \cref{tab:cost_heisenberg}, together with a breakdown of its subcircuits, which were detailed in \cref{tab:dicke-cost}. 

More specifically, the compact \PR{}/\PL{} implementation illustrated in \cref{fig:compact_heisenberg}, starts with preparing an unbalanced Dicke state on 6 qubits. Consequently, this preparation of $\DickeOne{6}(\alpha)$ requires only $5$ $\subr{\theta}$ subcircuits, resulting in a total of $10$ \cnots{}.
Next, we append:
two $\subr{\theta}$ subcircuits, two $\staircase{\n-1}$ subcircuits, two controlled $\CChain{1}$ subcircuits, one controlled $\copygate{}$ subcircuit,
and ten additional \cnots{}.
Controlling the $\CChain{1}$ and $\copygate{}$ subcircuits introduces the only Toffoli gates in the circuit, with a total count of $3\n-2$ for both $\PR$ and $\PL$. These, together with the $4\n+16$ \cnots{} from the other subroutines, yield a total cost of $22\n+4$ \cnots{} when each Toffoli is decomposed into $6$ \cnot{} gates \cite{Nielsen_2011}.

Finally, the \select{} oracle contributes with only $\n$ \cnot{} and $\n$ \cz{} gates, leading to a total cost of $46\n+8$ \cnots{} for the complete Heisenberg \cmlcu{} block encoding.

\begin{table}
\centering
\begin{tabular}{|c|c|c|c|}
\hline
\multicolumn{4}{|c|}{\textbf{Heisenberg}} \\
\hline
\textbf{Component} & \textbf{Subcircuits} & \textbf{Toffoli} & \textbf{\cnots{}} \\
\hline
\multirow{6}{*}{$\PR$} 
& $\DickeOne{6}(\alpha)$ &  & $10$ \\
& $2\subr{\theta_\n}$ &  & $4$ \\
& $2\staircase{\n-1}$ &  & $4\n-8$ \\
& $2 \text{ controlled } \CChain{1}$ & $2\n-2$ & $12\n-12$ \\
& $1 \text{ controlled } \copygate{}$ & $\n$ & $6\n$ \\
& Extra \cnots{} & & 10 \\
\hline
\multirow{2}{*}{$\select{}$} & $\n\cnot{}$ & & $\n$ \\
& $\n\cz$ &  & $\n$ \\
\hline
\PL & same as \PR & $3n - 2$ & $22n + 4$ \\
\hline
\hline
\bf\cmlcu{} & \text{\bf\PR{}\,+\,\select{}\,+\,\PL{}} & $\boldsymbol{6\n-4}$ & $\boldsymbol{46\n+8}$  \\
\hline
\end{tabular}
\caption{Breakdown of the computational cost for implementing the \cmlcu{} block encoding for the Heisenberg model. The \cnot{} counts in the last column include the decomposition of each Toffoli into $6$ \cnots{} \cite{Nielsen_2011}.}
\label{tab:cost_heisenberg}
\end{table}

\subsection{Spin glass \cmlcu{} block encoding}
\label{section:spin_glass_cost}
For the spin glass model, we analyze the simplified $\PR$ oracle described in \cref{subsec: compression Spin Glass}.
The total cost of the spin glass \cmlcu{} block encoding is summarized in \cref{tab:cost_spin_glass}.

\begin{table*}
\centering
\begin{tabular}{|c|c|c|c|}
\hline
\multicolumn{4}{|c|}{\textbf{Spin glass}} \\
\hline
\textbf{Component} & \textbf{Subcircuits} & \textbf{Toffoli} & \textbf{\cnots{}} \\
\hline
\multirow{6}{*}{$\PR$} 
& \DickeOneU{3\n}{\pscoeff} &  & $6\n-2$ \\
& $(3\n^2-3\n)\text{ controlled }\Rzempty$ &  & $6\n^2-6\n$ \\
& $2\text{ controlled }\CChain{k}\text{ for }\kN\in\{1,\ldots,\n-1\}$ & $\n^2-\n$ & $6\n^2-6\n$\\
& $1\text{ controlled }\copygate{}$ & $\n$ & $6\n$ \\
& Extra \cnots{} & & $11\n-8$ \\
& max $ \frac{3}{2}(\n^2+\n)\text{ controlled }P_\eta$ &  & max $ 3\n^2+3\n$ \\
\hline
\multirow{2}{*}{$\select{}$} & $\n\cnot{}$ & & $\n$ \\
& $\n\cz$ &  & $\n$ \\
\hline
$\PL{}$ & same as $\Pr{}$ & $\n^2$ & \makecell[c]{$12\n^2+11\n-10$ \\ to $15\n^2+14\n-10$} \\     
\hline
\hline
\bf\cmlcu{} & {\bf\PR{} + \select{} + \PL{}} & $\boldsymbol{2\n^2}$ & \makecell[c]{$\boldsymbol{24\n^2 + 24\n - 20}$ \\ to $\boldsymbol{30\n^2 + 30\n - 20}$}  \\
\hline
\end{tabular}
\caption{Breakdown of the computational cost for implementing the \cmlcu{} block encoding for the spin glass model. The \cnot{} counts in the last column include the decomposition of each Toffoli in $6$ \cnots{}, each controlled $\Rzempty$ in $2$ \cnots{} \cite{Nielsen_2011}, and each controlled $P_\eta$ in $2$ \cnots{} as shown in \cref{eq_decomposition_cphase}.}
\label{tab:cost_spin_glass}
\end{table*}

First of all, the \subPR{} and \subPL{} subcircuits act on $3\n$ qubits to prepare an unbalanced Dicke state.
As a next step, at least intuitively, each qubit of the additional $3\n$ ancillae prepares a Dicke state across the $\Xp$ and $\Zp$ ancilla registers, as initially illustrated in \cref{fig:spin_glass_prep}. However, after applying the simplifications described in \cref{sec: Compression controlled gates}, the individual oracles are no longer clearly distinguishable, and the resulting decomposition involves several components. 

First, there are \( 3\n \) \cnot{} gates used to activate the preparation of the Dicke states.
Then, for each direction \( x, y, z \) and for each value of \( \kN = 0, \ldots, \n - 1 \), there are \( 2(\n - \kN - 1) \) controlled-\( R_z \) gates, leading to a total of
\begin{equation}
    6 \cdot \sum_{\kN = 0}^{\n - 1} (\n - \kN - 1) = 3\n^2 - 3\n
\end{equation}
controlled-$R_z$ gates.
An additional $4\n - 4 $ \cnot{} gates are required due to the compression of the $\subr{\theta}$ subroutines, as described in \cref{eq:compression_gamma_spin_glass}.
A controlled phase gate $P_\eta$ is then appended for every negative coefficient appearing in \cref{eq:def_ham_spin_glass}, totalling up in the worst case scenario to:
\begin{equation}
    3 \sum_{\ell=1}^{\n} \ell = \tfrac{3}{2}(\n^2 + \n).
\end{equation}
Furthermore, for each $\kN \in \{1, \ldots, \n - 1\}$, two controlled $\CChain{\kN}$ gates are applied, which correspond to:
\begin{equation}
    2 \sum_{\kN = 1}^{\n - 1} (\n - \kN) = \n(\n - 1)
\end{equation}
Toffoli gates, along with an additional $2(\n - 1)$ \cnot{} gates, as derived in \cref{eq:compression_CLk}.
Finally, one controlled \copygate{} contributes to $\n$ Toffoli gates and an additional $2(\n - 1)$ \cnot{} gates, as shown in \cref{eq:simplification 11}.

Note that a controlled $P_\eta$ gate can be implemented with only $2$ \cnots{}:
\begin{equation}\label{eq_decomposition_cphase}
\begin{myqcircuit}
     & \ctrl{1} &\qw\\
     &\gate{P_{\eta}} &\qw
\end{myqcircuit}\; = \;
\begin{myqcircuit}
    &\qw &\ctrl{1} &\gate{P_{\frac{\eta}{2}}} &\ctrl{1}&\qw \\
    & \gate{P_{\frac{\eta}{2}}}&\targ &\gate{P_{-\frac{\eta}{2}}}&\targ &\qw
\end{myqcircuit}
\end{equation}
As a result, and considering $2$ \cnots{} for each controlled $R_z$ and $6$ \cnots{} for each Toffoli \cite{Nielsen_2011}, we get a total number of \cnots{} ranging between $24\n^2 + 24\n - 20 $ and $ 30\n^2 + 30\n - 20$.

\subsection{Benchmarking}
\label{section:comparison}
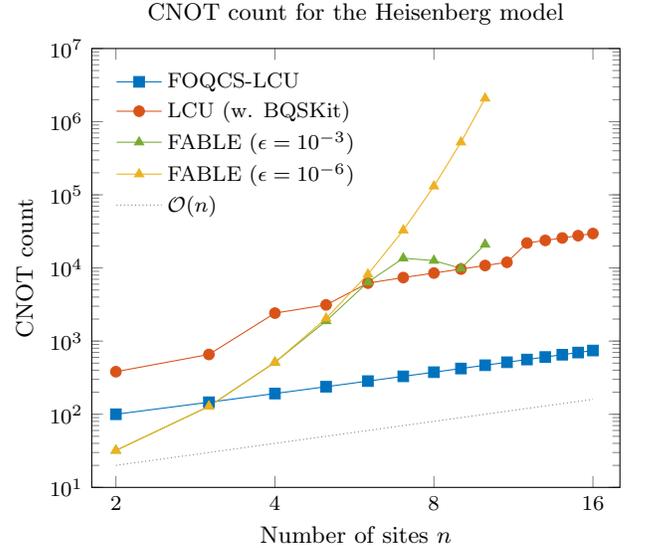
\begin{figure}
    \centering
    \begin{tikzpicture}
\begin{loglogaxis}[
  width=0.48\textwidth,
  log basis x=2,
  xmin=1.8, xmax=18,
  ymin=1e1, ymax=1e7,
  xtick={2,4,8,16},
  xticklabels={2,4,8,16},
  xlabel={Number of sites $\n$},
  ylabel={\cnot{} count},
  legend pos=north west,
  legend style={draw=none,fill=none},
  title={\cnot{} count for the Heisenberg model}
]

\addplot[myblue, mark=square*] table[x index=0, y index=1] {graphs/cmlcu_cnot.dat};

\addplot[myred, mark=*] table[x index=0, y index=1] {graphs/lcu_cnot.dat};

\addplot[mygreen, mark=triangle*] table[x index=0, y index=1] {graphs/fable3_cnot.dat};

\addplot[myorange,  mark=triangle*] table[x index=0, y index=1] {graphs/fable6_cnot.dat};

\addplot[no marks,gray,densely dotted]  table[x index=0, y index=1] {graphs/o_n.dat};


\legend{\cmlcu, LCU (w. BQSKit), FABLE ($\epsilon=10^{-3}$), FABLE ($\epsilon=10^{-6}$), $\bigO{\n}$} 
\end{loglogaxis}
\end{tikzpicture}
    \caption{\cnot{} count from \cmlcu, LCU, and FABLE for the Heisenberg model, with FABLE evaluated at precisions $\epsilon = 10^{-3}$ and $\epsilon = 10^{-6}$.}
    \label{fig:cnot_graph_heisenberg}
\end{figure}

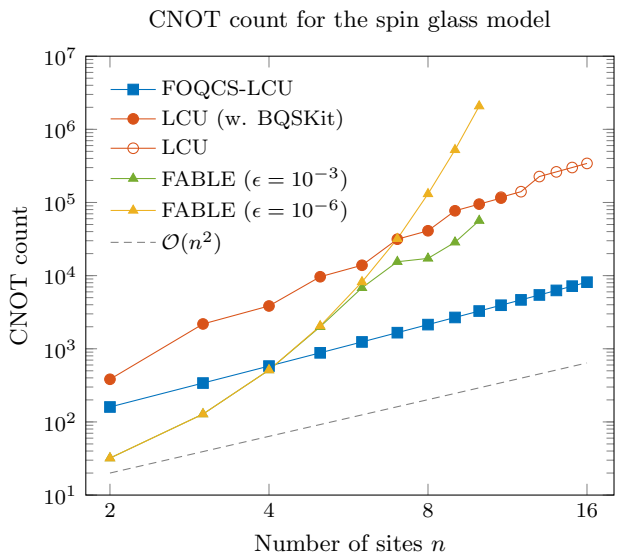
\begin{figure}
    \centering
    \begin{tikzpicture}
\begin{axis}[
  width=0.48\textwidth,
  xmode=log,
  log basis x=2,
  ymode=log,
  xmin=1.8, xmax=18,
  ymin=1e1, ymax=1e7,
  xtick={2,4,8,16},
  xticklabels={2,4,8,16},
  xlabel={Number of sites $\n$},
  ylabel={\cnot{} count},
  legend pos=north west,
  legend style={draw=none,fill=none},
  title={\cnot{} count for the spin glass model}
]

\addplot[myblue, mark=square*] table[x index=0, y index=2] {graphs/cmlcu_cnot_sg.dat};

\addplot[myred, mark=*] table[x index=0, y index=1] {graphs/lcu_cnot_sg_bqskit.dat};
\addplot[myred, mark=o] table[x index=0, y index=1] {graphs/lcu_cnot_sg.dat};

\addplot[mygreen, mark=triangle*] table[x index=0, y index=1] {graphs/fable3_cnot_sg.dat};

\addplot[myorange,  mark=triangle*] table[x index=0, y index=1] {graphs/fable6_cnot_sg.dat};

\addplot[no marks,gray,densely dashed]  table[x index=0, y index=1] {graphs/o_n2.dat};

\legend{\cmlcu, LCU (w. BQSKit), LCU, FABLE ($\epsilon=10^{-3}$), FABLE ($\epsilon=10^{-6}$), $\bigO{\n^2}$} 
\end{axis}
\end{tikzpicture}
    \caption{\cnot{} count from \cmlcu, LCU, and FABLE for the spin glass model, with FABLE evaluated at precisions $\epsilon = 10^{-3}$ and $\epsilon = 10^{-6}$. For \cmlcu, we considered the worst-case scenario in which $\frac{3}{2}(\n^2+\n)$ controlled phase gates are needed.}
    \label{fig:cnot_graph_spin_glass}
\end{figure}

In \cref{fig:cnot_graph_heisenberg} and \cref{fig:cnot_graph_spin_glass}, we compare the \cnot{} counts of \cmlcu{} implementations for the Heisenberg and spin glass models, respectively, with the standard LCU method. 
All circuits were generated through the \texttt{qiskit} software \cite{qiskit2024}, with the source code available at \url{https://github.com/QuantumComputingLab/foqcs-lcu}. A Matlab implementation of the Heisenberg model through the \texttt{QCLab} software \cite{Keip2025} is also available.
Both figures confirm the theoretical scaling of the structure-aware \cmlcu{}, which scales as $\bigO{\n}$ for the Heisenberg model and as $\bigO{\n^2}$ for the spin glass model. For the spin glass model, we plotted the worst-case scenario in which additional $\frac{3}{2}(\n^2+\n)$ controlled phase gates are needed.

By contrast, the standard LCU circuit shown in \cref{fig:lcu_general} does not take any specific structure of the model under consideration into account in its implementation. As a result, it exhibits poorer scaling compared to our \cmlcu{} method, which explicitly leverages the problem structure in its design. The \PR{} and \PL{} subcircuits in \cref{fig:lcu_general}, defined in \cref{eq:def_PR,eq:def_PL}, were instantiated using \texttt{qiskit}’s default state initialization routine \cite{Iten2016}. To reduce circuit depth and cancel redundant gates, we applied post-compilation optimization with the Berkeley Quantum Synthesis Toolkit (BQSKit) \cite{bqskit2021}, which implements state-of-the-art synthesis techniques.
BQSKit was applied to all data points for the Heisenberg model. However, for the spin glass Hamiltonian, we could only apply BQSKit up to 10 spin sites, as the circuit already involves 60 qubits, exceeding the practical limits of the synthesis tool. Despite this additional optimization step, BQSKit provides only modest improvements (about 10\% gate reduction), without affecting the theoretical scaling of the standard LCU approach.

As illustrated in \cref{fig:cnot_graph_heisenberg,fig:cnot_graph_spin_glass}, \cmlcu{} exhibits better scaling than the standard, structureless LCU and achieves over an order of magnitude reduction in circuit size for both models within the considered spin-chain sizes.
We also benchmark against the Fast Approximate Block Encodings (FABLE) algorithm \cite{Camps2022}, evaluated at two different accuracy thresholds. FABLE is an explicit, approximate, and general-purpose block encoding method that formally applies to any matrix, with a worst-case cost that scales exponentially. However, it is particularly effective for sparse matrices, where this cost can be drastically mitigated. In our setting, the sparsity of the Hamiltonians makes FABLE a valid candidate for numerical benchmarking. As shown in \cref{fig:cnot_graph_heisenberg,fig:cnot_graph_spin_glass}, FABLE can indeed be competitive with LCU (in both its standard and FOQCS variants) for small system sizes. However, its exponential scaling quickly makes it unsuitable for larger system sizes.

\section{Conclusion}
\label{section:conclusion}

In this work, we have introduced a practical, low-depth formulation of the LCU block encoding, with explicit circuits for a variety of spin models. By exploiting the structure of the underlying Hamiltonians, our \cmlcu{} implementations achieve over an order-of-magnitude reduction in total \cnot{} count compared to the standard, structure-agnostic LCU. This improvement arises from two key ingredients: $(1)$ the design of structured state-preparation routines based on generalized Dicke states, and $(2)$ the replacement of multi-controlled gates in the \select{} oracle with two fully parallelizable layers of $\n$ singly controlled Pauli gates.

More specifically, both circuit simplifications rely on allocating more ancilla qubits. This trade-off between qubit count and circuit depth can be particularly favorable in the current stage of quantum hardware development, as we transition from the late NISQ era into early fault-tolerant quantum computing, where qubit count is becoming less of a constraint compared to the circuit depth and noise resilience. Notably, recent work proposes strategies for reducing ancilla overhead in general block encodings~\cite{vasconcelos2025}.
This could further improve our constructions in the fault-tolerant regime.

Our approach naturally generalizes beyond the spin chains studied here. We outline how state preparation oracles can be extended to accommodate arbitrary local interactions.
Moreover, recent developments suggest that combining Dicke state constructions with mid-circuit measurements enables constant-depth implementations in certain contexts \cite{Piroli2024, Buhrman2024, Farrell2025}. An open question is whether similar techniques can be adapted to our framework, which requires controlled versions of these subcircuits, to further reduce the total circuit depth in the \cmlcu{} algorithm.

\begin{acknowledgments}
This research was supported by the U.S. Department of Energy (DOE) under Contract No.~DE-AC02-05CH11231, through the Office of Science, Office of Advanced Scientific Computing Research (ASCR), Accelerated Research in Quantum Computing.
MN acknowledges funding by the Munich Quantum Valley, section K5 Q-DESSI. The research is part of the Munich Quantum Valley, which is supported by the Bavarian state government with funds from the Hightech Agenda Bayern Plus.
\end{acknowledgments}

\bibliography{bibliography}

\newpage
\onecolumngrid
\appendix
\labelformat{subsection}{\thesection#1}

\section{Proof of the correctness of the \cmlcu{} circuits}
\label{section:proof_general}

\subsection{Block encoding of a 1-qubit matrix}
\label{sec:proof-1q}

We prove that the circuit in \cref{eq:be_general_simple} correctly prepares the matrix $\generalmat$ of dimension $2\times 2$ from \cref{eq:lcu_simple_1} and \cref{eq:lcu_simple_2}.
Given a starting one-qubit state $\ket{\varphi}$ on which $\generalmat$ is applied:
\begin{equation}
    \generalmat\ket{\varphi} = \left(\pscoeff_{00} \Ip + \pscoeff_{01}\Zp + \pscoeff_{10}\Xp -i\pscoeff_{11}\Zp\Xp\right) \ket{\varphi},
\end{equation}
the circuit yields the following outcomes at each step:
\begin{align}
\ket{00}\ket{\varphi} \overset{\PR}{\longrightarrow}& \left(\sqrt{\pscoeff_{00}}\ket{00}+\sqrt{\pscoeff_{01}}\ket{01}+\sqrt{\pscoeff_{10}}\ket{10}+\sqrt{-i\pscoeff_{11}}\ket{11}\right)\ket{\varphi}, \\
\overset{\cnot}{\longrightarrow}&  \sqrt{\pscoeff_{00}}\ket{00}\ket{\varphi}+\sqrt{\pscoeff_{01}}\ket{01}\ket{\varphi}+\sqrt{\pscoeff_{10}}\ket{10}\Xp\ket{\varphi}+\sqrt{-i\pscoeff_{11}}\ket{11}\Xp\ket{\varphi}, \\
\overset{\cz}{\longrightarrow}& \sqrt{\pscoeff_{00}}\ket{00}\ket{\varphi}+\sqrt{\pscoeff_{01}}\ket{01}\Zp\ket{\varphi}+\sqrt{\pscoeff_{10}}\ket{10}\Xp\ket{\varphi}+\sqrt{-i\pscoeff_{11}}\ket{11}\Zp\Xp\ket{\varphi}.
\label{eq:app-proof-1q-A}
\end{align}
Before measuring the ancillae, we apply the gate $\PLdag$, whose action is defined in \cref{eq:def_prep_1q_PL}.
Specifically, by applying its conjugate transpose, we obtain:
\begin{equation}
\bra{00}\PLdag =\sqrt{\pscoeff_{00}}\bra{00}+\sqrt{\pscoeff_{01}}\bra{01}+\sqrt{\pscoeff_{10}}\bra{10}+\sqrt{-i\pscoeff_{11}}\bra{11}.
\label{eq:app-proof-1q-B}
\end{equation}
Finally, taking the inner product of \cref{eq:app-proof-1q-A,eq:app-proof-1q-B} yields:
\begin{equation}
\bra{00}\PLdag\cdot\select\cdot\PR\ket{00}\ket{\varphi} =
\pscoeff_{00}\ket{\varphi}+\pscoeff_{01}\Zp\ket{\varphi}+\pscoeff_{10}\Xp\ket{\varphi}-i\pscoeff_{11}\Zp\Xp\ket{\varphi} = \generalmat\ket{\varphi}.
\end{equation}

\subsection{Block encoding of an $n$-qubit matrix}
\label{sec:proof-nq}

For the $\n$ qubits case, on the other hand, we showed in \cref{section:main_result} that it's always possible to decompose a generic matrix $\generalmat$ of dimension $2^\n \times 2^\n$ as in \cref{eq:lcu_general_check_matrix}.

We now want to prove that the circuit in \cref{fig:be_general}, with $\PR$ defined in \cref{eq:PR_state_general}, yields $\generalmat\ket{\varphi}$, where $\ket{\varphi}$ is a generic $\n$-qubit state.
In order to do that, at first we fix the two $\n$-qubit ancillary registers as $\ket{\ix} = \ket{\ixbit{\n-1},\dots,\ixbit{0}}$ and $\ket{\jz} = \ket{\jzbit{\n-1},\dots,\jzbit{0}}$.

Applying the $\n$ $\cnots$ of the $\select$ oracle will result in the following state:
\begin{equation}
\left(\Xp^{\ixbit{\n-1}}\otimes \Xp^{\ixbit{\n-2}}\otimes \ldots \otimes \Xp^{i_{0}} \right) \ket{\ixbit{\n-1},\dots,\ixbit{0}}\ket{\jzbit{\n-1},\dots,\jzbit{0}}\ket{\varphi} \; .
\end{equation}
Then, after the controlled $\Zp$, we get:
\begin{equation}
\longrightarrow\left(\Zp^{\ixbit{\n-1}}\Xp^{\ixbit{\n-1}}\otimes \Zp^{\jzbit{\n-2}}\Xp^{\ixbit{\n-2}}\otimes \ldots \otimes \Zp^{\jzbit{0}}\Xp^{\ixbit{0}} \right)\ket{\ixbit{\n-1},\dots,\ixbit{0}}\ket{\jzbit{\n-1},\dots,\jzbit{0}}\ket{\varphi} =\bigotimes_{\elle=0}^{\n-1}\Zp^{\jzbit{\elle}}\Xp^{\ixbit{\elle}}\ket{\ix}\ket{\jz}\ket{\varphi}   \; ,
\end{equation}
which proves the result for fixed $\ket{\ix}$ and $\ket{\jz}$.

If we now consider the $\PR$ and $\PLdag$ gates, whose actions were defined respectively in \cref{eq:def_PR} and \cref{eq:def_PL}, we can now prove that the circuit in \cref{fig:be_general} implements the block encoding of $\generalmat$:
\begin{align}
\bra{0^{\otimes\n}}\bra{0^{\otimes\n}}\PLdag\cdot\select\cdot\PR\ket{0^{\otimes\n}} \ket{0^{\otimes\n}}\ket{\varphi} &= \left[\sum_{\ix^\prime=0}^{2^{\n-1}} \sum_{\jz^\prime=0}^{2^{\n-1}} \sqrt{\newpscoeff_{\ix^\prime\jz^\prime}} \bra{\ix^\prime}\bra{\jz^\prime}\right] \select \left[\sum_{\ix=0}^{2^{\n-1}} \sum_{\jz=0}^{2^{\n-1}} \sqrt{\newpscoeff_{\ix\jz}} \ket{\ix}\ket{\jz}\right]\ket{\varphi}\nonumber \\
    &= \left[\sum_{\ix^\prime=0}^{2^{\n-1}} \sum_{\jz^\prime=0}^{2^{\n-1}} \sqrt{\newpscoeff_{\ix^\prime\jz^\prime}} \bra{\ix^\prime}\bra{\jz^\prime}\right] \left[\sum_{\ix=0}^{2^{\n-1}} \sum_{\jz=0}^{2^{\n-1}} \sqrt{\newpscoeff_{\ix\jz}} \ket{\ix}\ket{\jz} \bigotimes_{\elle=0}^{\n-1}\Zp^{\jzbit{\elle}}\Xp^{\ixbit{\elle}}\ket{\varphi} \right]  \nonumber \\
    &= \sum_{\ix=0}^{2^{\n-1}} \sum_{\jz=0}^{2^{\n-1}} \newpscoeff_{\ix\jz} \bigotimes_{\elle=0}^{{\n-1}}\Zp^{\jzbit{\elle}}\Xp^{\ixbit{\elle}}\ket{\varphi} \nonumber \\[1em]
    &= \generalmat\ket{\varphi}
\end{align}

\section{Preparation of unbalanced Dicke states with $\excitations=1$}
\label{section:unbalanced_dicke}

In this section, we show that the circuit in \cref{eq_dicke_unbalanced_oracle} correctly prepares the unbalanced Dicke state defined in \cref{eq:dicke one unbalanced def}, where the subcircuit $\subr{\hat{\theta}}$ is defined as \cref{fig:subr_gamma} and the angles $\hat{\theta}$ are specified by \cref{eq_theta_unbalanced_general,eq_theta_unbalanced_n-1}.
Recall that the phase gates coefficients $\coeffDicke{\elle}=\vert\coeffDicke{\ell}\vert e^{i\eta_\elle}$ are generally complex.

We start by proving that the first part of the circuit (before applying the phase gates $P_{\eta_\elle}$) correctly prepares the state in \cref{eq:dicke one unbalanced def} up to some relative phases, which are corrected by the application of the $P_{\eta_\elle}$ gates.
First of all, it is important to observe the outcome of the application of $\subr{\hat\theta_{\n-1}}$ on the initial two-qubit state $\ket{01}$:
\begin{align}
    \ket{\cdots01} \xrightarrow{\subr{\hat\theta_{\n-1}}}& \; \cos\frac{\hat\theta_{\n-1}}{2} \ket{\cdots01} + \sin\frac{\hat\theta_{\n-1}}{2} \ket{\cdots10} \nonumber\\
    =& \; \vert\coeffDicke{0}\vert\ket{2^0} + \sqrt{1-\vert\coeffDicke{0}\vert^2}\ket{2^{1}} \; .
\end{align}

Subsequently, we get at the second step:
\begin{align}
\label{eq:second_step_dicke_unbalanced_proof}
\xrightarrow{\subr{\hat\theta_{\n-2}}}& \; \vert\coeffDicke{0}\vert\ket{\cdots001} + \cos\frac{\hat\theta_{\n-2}}{2}  \sqrt{1-\vert\coeffDicke{0}\vert^2}\ket{\cdots010} + \sin \frac{\hat\theta_{\n-2}}{2} \sqrt{1-\vert\coeffDicke{0}\vert^2}\ket{\cdots100}  \nonumber\\
=& \; \vert\coeffDicke{0}\vert\ket{2^{0}} + \vert\coeffDicke{1}\vert\ket{2^{1}} + \sqrt{1-\vert\coeffDicke{1}\vert^2 - \vert\coeffDicke{0}\vert^2} \ket{2^{2}} \; ,
\end{align}
where we have used the following relations:
\begin{align}
    \cos \frac{\hat\theta_{\n-2}}{2} &= \frac{\vert \coeffDicke{1}\vert}{\sqrt{1-\vert\coeffDicke{0}\vert^2\ }}\; , \\
\sin\frac{\hat\theta_{\n-2}}{2} &= \sqrt{1-\left(\cos\frac{\hat\theta_{\n-2}}{2}\right)^2} = \sqrt{1-\frac{\vert \coeffDicke{1}\vert^2}{1-\vert\coeffDicke{0}\vert^2 }} = \frac{\sqrt{1-\vert \coeffDicke{1}\vert^2 -\vert\coeffDicke{0}\vert^2}}{\sqrt{1-\vert\coeffDicke{0}\vert^2}} \; .
\end{align}

In a recursive manner, we can then extract the final state:
\begin{equation} 
\xrightarrow{\staircase{\n}(\hat{\theta})} \sum_{\elle=0}^{\n-2} \vert\coeffDicke{\elle}\vert\ket{2^\elle} + \sqrt{1-\sum_{\elle=0}^{\n-2} \vert \coeffDicke{\elle}\vert^2} \ket{2^{\n-1}} \; ,
\end{equation}
where it is sufficient to observe that:
\begin{equation} 
\sqrt{1-\sum_{\elle=0}^{\n-2} \vert \coeffDicke{\elle}\vert^2} = \vert\coeffDicke{\n-1}\vert \; ,
\end{equation}
since $\Vert\coeffDicke{}\Vert_2 = 1 $, to get the state in \cref{eq:dicke one unbalanced def} up to some relative phases.

Finally, in order to correct the complex phases $\eta_\elle$, it is sufficient to apply the phase gates defined in \cref{eq:def_phase_gate} on each qubit, as shown in \cref{eq_dicke_unbalanced_oracle}.

\section{2 \cnot{} implementation for $\subr{\theta}$}
\label{section: proof of 2 cnot}

Here, we present an implementation of the $\subr{\theta}$ subroutine that requires only two \cnots{}. The circuit is given below:
\begin{equation}
    \begin{myqcircuit}
    & \gate{S} &\gate{H} &\cgate{R_z(\frac\pi2-\frac\theta2)}{Rzcolor} &\ctrl{1} &\qw &\gate{H} &\qw &\ctrl{1} &\gate{S} &\gate{H}&\qw\\
    &\qw &\qw &\qw &\targ{} &\cgate{R_z(\frac\theta2-\frac\pi2)}{Rzcolor}&\gate{H}&\gate{\conjugate{S}} &\targ{} & \qw &\qw &\qw
    \end{myqcircuit} 
    \label{eq: 2cnot subr}
\end{equation}
Then we just prove that this circuit will implement the following for the basis states:
\begin{itemize}
    \item $\ket{00}\longrightarrow \ket{00}$
    \item $\ket{01} \longrightarrow \cos\frac\theta2 \ket{01} + \sin\frac\theta2\ket{10}$
    \item $\ket{10} \longrightarrow \ket{11}$
    \item $\ket{11} \longrightarrow -\sin\frac\theta2 \ket{01} +\cos\frac\theta2 \ket{10}$
\end{itemize}
We now show that the two circuits implement the same transformation on computational basis states.
Starting with $\ket{00}$, we have:
\begin{alignat}{3}
    \nonumber\ket{00}&\xrightarrow {\left( R_z(\frac\pi2-\frac\theta2)\cdot H\cdot S\right)\otimes I_2} \; &&\frac{e^{-i(\frac\pi4-\frac\theta4)}}{\sqrt2} \ket{00}+\frac{e^{i(\frac\pi4-\frac\theta4)}}{\sqrt2} \ket{10}\\
    \nonumber&\xrightarrow{\cnot{}} &&\frac{e^{-i(\frac\pi4-\frac\theta4)}}{\sqrt2} \ket{00}+\frac{e^{i(\frac\pi4-\frac\theta4)}}{\sqrt2} \ket{11}\\
    \nonumber& \xrightarrow{I_2\otimes R_z(\frac\theta2-\frac\pi2)} &&e^{-i(\frac\theta4-\frac\pi4)}\cdot\frac{e^{-i(\frac\pi4-\frac\theta4)}}{\sqrt2} \ket{00}+e^{i(\frac\theta4-\frac\pi4)}\cdot\frac{e^{i(\frac\pi4-\frac\theta4)}}{\sqrt2} \ket{11}\, ,
\end{alignat}
where $e^{-i(\frac\theta4-\frac\pi4)}\cdot\frac{e^{-i(\frac\pi4-\frac\theta4)}}{\sqrt2} \ket{00}+e^{i(\frac\theta4-\frac\pi4)}\cdot\frac{e^{i(\frac\pi4-\frac\theta4)}}{\sqrt2} \ket{11} = \frac{1}{\sqrt2}\ket{00} + \frac{1}{\sqrt2}\ket{11}$, so that:
\begin{alignat}{3}
    \frac{1}{\sqrt2}\ket{00} + \frac{1}{\sqrt2}\ket{11}
    \nonumber& \xrightarrow{H\otimes H}  \frac{1}{\sqrt2}\ket{00} + \frac{1}{\sqrt2}\ket{11} \\
    \nonumber&\xrightarrow{I_2\otimes \conjugate{S}}  \frac{1}{\sqrt2}\ket{00} - \frac{i}{\sqrt2}\ket{11}\\
    \nonumber&\xrightarrow{\cnot{}}  \frac{1}{\sqrt2}\ket{00} - \frac{i}{\sqrt2}\ket{10}\\
    \nonumber&\xrightarrow{S\otimes I_2}  \frac{1}{\sqrt2}\ket{00} + \frac{1}{\sqrt2}\ket{10}\\
    &\xrightarrow{H\otimes I_2}  \ket{00} \; .
\end{alignat}
For $\ket{10}$, we have:
\begin{alignat}{3}
    \nonumber\ket{10}&\xrightarrow {\left( R_z(\frac\pi2-\frac\theta2)\cdot H\cdot S\right)\otimes I_2} \;&&i\frac{e^{-i(\frac\pi4-\frac\theta4)}}{\sqrt2} \ket{00}-i\frac{e^{i(\frac\pi4-\frac\theta4)}}{\sqrt2} \ket{10}\\
    \nonumber&\xrightarrow{\cnot{}} &&i\frac{e^{-i(\frac\pi4-\frac\theta4)}}{\sqrt2} \ket{00}-i\frac{e^{i(\frac\pi4-\frac\theta4)}}{\sqrt2} \ket{11}\\
    \nonumber& \xrightarrow{I_2\otimes R_z(\frac\theta2-\frac\pi2)} &&e^{-i(\frac\theta4-\frac\pi4)}\cdot\frac{i\cdot e^{-i(\frac\pi4-\frac\theta4)}}{\sqrt2} \ket{00}-e^{i(\frac\theta4-\frac\pi4)}\cdot\frac{i\cdot e^{i(\frac\pi4-\frac\theta4)}}{\sqrt2} \ket{11} \, ,
\end{alignat}
where $e^{-i(\frac\theta4-\frac\pi4)}\cdot\frac{i\cdot e^{-i(\frac\pi4-\frac\theta4)}}{\sqrt2} \ket{00}-e^{i(\frac\theta4-\frac\pi4)}\cdot\frac{i\cdot e^{i(\frac\pi4-\frac\theta4)}}{\sqrt2} \ket{11} = \frac{i}{\sqrt2}\ket{00} - \frac{i}{\sqrt2}\ket{11}$, so that:
\begin{alignat}{3}
    \frac{i}{\sqrt2}\ket{00} - \frac{i}{\sqrt2}\ket{11}
    \nonumber& \xrightarrow{H\otimes H}  \frac{i}{\sqrt2}\ket{01} + \frac{i}{\sqrt2}\ket{10} \\
    \nonumber&\xrightarrow{I_2\otimes \conjugate{S}}  \frac{1}{\sqrt2}\ket{01} - \frac{i}{\sqrt2}\ket{10}\\
    \nonumber&\xrightarrow{\cnot{}}  \frac{1}{\sqrt2}\ket{01} - \frac{i}{\sqrt2}\ket{11}\\
    \nonumber&\xrightarrow{S\otimes I_2}  \frac{1}{\sqrt2}\ket{01} - \frac{1}{\sqrt2}\ket{11}\\
    &\xrightarrow{H\otimes I_2}  \ket{11} \; .
\end{alignat}
For $\ket{01}$, we get: 
\begin{alignat}{3}
    \nonumber\ket{01}&\xrightarrow {\left( R_z(\frac\pi2-\frac\theta2)\cdot H\cdot S\right)\otimes I_2} &&\frac{e^{-i(\frac\pi4-\frac\theta4)}}{\sqrt2} \ket{01}+\frac{e^{i(\frac\pi4-\frac\theta4)}}{\sqrt2} \ket{11}\\
    \nonumber&\xrightarrow{\cnot{}} &&\frac{e^{-i(\frac\pi4-\frac\theta4)}}{\sqrt2} \ket{01}+\frac{e^{i(\frac\pi4-\frac\theta4)}}{\sqrt2} \ket{10}\\
    \nonumber& \xrightarrow{I_2\otimes R_z(\frac\theta2-\frac\pi2)} &&e^{i(\frac\theta4-\frac\pi4)}\cdot\frac{e^{-i(\frac\pi4-\frac\theta4)}}{\sqrt2} \ket{01}+e^{-i(\frac\theta4-\frac\pi4)}\cdot\frac{e^{i(\frac\pi4-\frac\theta4)}}{\sqrt2} \ket{10} \, ,
\end{alignat}
where we have that $e^{i(\frac\theta4-\frac\pi4)}\cdot\frac{e^{-i(\frac\pi4-\frac\theta4)}}{\sqrt2} \ket{01}+e^{-i(\frac\theta4-\frac\pi4)}\cdot\frac{e^{i(\frac\pi4-\frac\theta4)}}{\sqrt2} \ket{10} = \frac{-i\cdot e^{i\frac\theta2}}{\sqrt2}\ket{01} + \frac{i\cdot e^{-i\frac{\theta}{2}}}{\sqrt2}\ket{10}$, so that: 
\begin{alignat}{3}
    \frac{-i\cdot e^{i\frac\theta2}}{\sqrt2}\ket{01} + \frac{i\cdot e^{-i\frac{\theta}{2}}}{\sqrt2}\ket{10}
    \nonumber& \xrightarrow{H\otimes H} && \frac{1}{\sqrt{2}} \left(\sin\frac{\theta}{2}\ket{00}+ i\cos\frac{\theta}{2}\ket{01}- i\cos\frac{\theta}{2}\ket{10}- \sin\frac{\theta}{2}\ket{11}\right)\\
    \nonumber&\xrightarrow{I_2\otimes \conjugate{S}} && \frac{1}{\sqrt2}\left(\sin\frac\theta2\ket{00} + \cos\frac\theta2\ket{01} -i\cos\frac\theta2\ket{10} +i \sin\frac\theta2\ket{11}\right )\\
    \nonumber&\xrightarrow{\cnot{}} && \frac{1}{\sqrt2}\left(\sin\frac\theta2\ket{00} + \cos\frac\theta2\ket{01} -i\cos\frac\theta2\ket{11} +i \sin\frac\theta2\ket{10}\right )\\
    \nonumber&\xrightarrow{S\otimes I_2} && \frac{1}{\sqrt2}\left(\sin\frac\theta2\ket{00} + \cos\frac\theta2\ket{01} +\cos\frac\theta2\ket{11} - \sin\frac\theta2\ket{10}\right )\\
    &\xrightarrow{H\otimes I_2} && \cos\frac\theta2 \ket{01} + \sin\frac\theta2\ket{10} \; .
\end{alignat}
Finally, for $\ket{11}$ we have:
\begin{alignat}{3}
    \nonumber\ket{11}&\xrightarrow {\left( R_z(\frac\pi2-\frac\theta2)\cdot H\cdot S\right)\otimes I_2} &&i\frac{e^{-i(\frac\pi4-\frac\theta4)}}{\sqrt2} \ket{01}-i\frac{e^{i(\frac\pi4-\frac\theta4)}}{\sqrt2} \ket{11}\\
    \nonumber&\xrightarrow{\cnot{}} &&\frac{e^{-i(\frac\pi4-\frac\theta4)}}{\sqrt2} \ket{01}-\frac{e^{i(\frac\pi4-\frac\theta4)}}{\sqrt2} \ket{10}\\
    \nonumber& \xrightarrow{I_2\otimes R_z(\frac\theta2-\frac\pi2)} &&e^{i(\frac\theta4-\frac\pi4)}\cdot\frac{i\cdot e^{-i(\frac\pi4-\frac\theta4)}}{\sqrt2} \ket{01}-e^{-i(\frac\theta4-\frac\pi4)}\cdot\frac{i\cdot e^{i(\frac\pi4-\frac\theta4)}}{\sqrt2} \ket{10}\, ,
\end{alignat}
where $e^{i(\frac\theta4-\frac\pi4)}\cdot\frac{i\cdot e^{-i(\frac\pi4-\frac\theta4)}}{\sqrt2} \ket{01}-e^{-i(\frac\theta4-\frac\pi4)}\cdot\frac{i\cdot e^{i(\frac\pi4-\frac\theta4)}}{\sqrt2} \ket{10} = \frac{ e^{i\frac\theta2}}{\sqrt2}\ket{01} + \frac{ e^{-i\frac{\theta}{2}}}{\sqrt2}\ket{10}$, so that:
\begin{alignat}{3}
    \frac{ e^{i\frac\theta2}}{\sqrt2}\ket{01} + \frac{ e^{-i\frac{\theta}{2}}}{\sqrt2}\ket{10}
     \nonumber &\xrightarrow{H\otimes H} && \frac{1}{\sqrt{2}} \left( \cos\frac{\theta}{2} \ket{00} - i\sin\frac{\theta}{2} \ket{01}+ i\sin\frac{\theta}{2} \ket{10}- \cos\frac{\theta}{2} \ket{11} \right)\\
    \nonumber&\xrightarrow{I_2\otimes \conjugate{S}} && \frac{1}{\sqrt{2}} \left( \cos\frac{\theta}{2} \ket{00}  -\sin\frac{\theta}{2} \ket{01}+ i\sin\frac{\theta}{2} \ket{10}+i \cos\frac{\theta}{2} \ket{11} \right)\\
    \nonumber&\xrightarrow{\cnot{}} && \frac{1}{\sqrt{2}} \left( \cos\frac{\theta}{2} \ket{00}  -\sin\frac{\theta}{2} \ket{01}+ i\sin\frac{\theta}{2} \ket{11}+i \cos\frac{\theta}{2} \ket{10} \right)\\
    \nonumber&\xrightarrow{S\otimes I_2} && \frac{1}{\sqrt{2}} \left( \cos\frac{\theta}{2} \ket{00}  -\sin\frac{\theta}{2} \ket{01}-\sin\frac{\theta}{2} \ket{11}- \cos\frac{\theta}{2} \ket{10} \right)\\
    &\xrightarrow{H\otimes I_2} && -\sin\frac\theta2 \ket{01} + \cos\frac\theta2\ket{10} \;.
\end{alignat}
This concludes the proof.

\section{Compression of $\PR$ for the Heisenberg and Spin Glass Hamiltonians}
\label{sec: Compression controlled gates}

In this section, we focus on simplifying the preparation step for the Hamiltonians introduced in \cref{section:applications}. The proposed optimizations exploit the fact that many of the controlled subroutines in $\PR$ and $\PL$ are repeated, even if with a different control qubit. 
Since, as defined in \cref{section:applications}, the state $\ket{\subPR}$ is an unbalanced Dicke state with $\excitations=1$, many of these controls can be removed, leading to a significant reduction in the number of \cnot{} gates.

In \cref{subsec: controlled same gates}, we begin by showing how to contract two identical controlled gates acting on the same target but with different controls. This technique is key to reducing the number of applications of the subroutines \CChain{\kN} and \copygate{}.
Subsequently, in \cref{subsec: controlled balanced Dicke} and \cref{subsec: controlled unbalanced Dicke}, we demonstrate how to contract the controlled subroutines $\subr{\theta}$ that appear in the preparation of balanced and unbalanced Dicke states, respectively.
These protocols are then applied in \cref{subsec: compression Heisenberg} and \cref{subsec: compression Spin Glass}, where we optimize the implementations of $\PR$ and $\PL$ for Heisenberg and spin glass models.

\subsection{Reduction of controlled \CChain{\kN} and \copygate{}}
\label{subsec: controlled same gates}
We begin by considering the following state equality:
\begin{equation}\label{eq:simplification 9}
\begin{myqcircuit}[0.1]
& \cdots && \\
& \qw & \ctrl{3} &\qw &\qw \\
& \qw &\qw &\ctrl{2}&\qw \\
& \cdots && \\
\lstick{\ket{\varphi}} &\qw&\gate{C} & \gate{C} &\qw \inputgroupv{1}{4}{1em}{2em}{\hspace{-0.7cm}\ket{\DickeOneU{\n}{\pscoeff}}} \\
\end{myqcircuit}
\;=\qquad\qquad\qquad\quad
\begin{myqcircuit}[0.1]
& \cdots && \\
 & \qw & \ctrl{1} &\qw &\ctrl{1}&\qw \\
 & \qw & \targ{}&\ctrl{2} &\targ{} &\qw\\
 & \cdots && \\
 \lstick{\ket{\varphi}}& \qw & \qw &\gate{C} & \qw & \qw & \qw  \inputgroupv{1}{4}{1em}{2em}{\hspace{-0.7cm}\ket{\DickeOneU{\n}{\pscoeff}}} \\
\end{myqcircuit}
\end{equation}
where $C$ is a generic gate and the controlling qubits are part of an unbalanced Dicke state with $\nu=1$, so that the states of the controls have at most one excitation.

Then, if we consider the circuits in \cref{fig:heisenberg_prep,fig:spin_glass_prep}, one observes that several subroutines are applied multiple times to the same target, but with different control qubits. This is the case, for example, for the $\CChain{\kN}$ subroutine, which is used in both the $\DickeTwokN{\n}{\kN}$ and $\DickeTwokND{\n}{\kN}$ protocols, as well as the \copygate{} subroutine, which duplicates a Dicke state.

Since \subPR{} prepares an unbalanced Dicke state with Hamming weight equal to $1$, it's possible to generalize \cref{eq:simplification 9} to reduce the number of applications of $\CChain{\kN}$ and $\copygate{}$.
More specifically, for every $\kN$ we will need only $2$ \CChain{\kN}, one for each of the \Xp{} and \Zp{} registers, instead of $3$:
\begin{equation} \label{eq:compression_CLk}
    \begin{myqcircuit}[0.1]
        & \cdots &&& \\
        & \qw & \qw & \ctrl{4} & \qw & \qw & \qw \\
        & \qw & \qw &\qw & \ctrl{4} & \qw & \qw \\
        & \qw & \qw &\qw & \qw & \ctrl{2} & \qw \inputgroupv{1}{5}{1em}{3em}{\hspace{-0.7cm}\ket{\DickeOneU{\n}{\pscoeff}}}\\
        & \cdots &&& \\
        & \qw {/\strut^{\n}} & \qw &\cgate{\CChain{\kN}}{CLkcolor} & \qw & \cgate{\CChain{\kN}}{CLkcolor} & \qw \\
        & \qw {/\strut^{\n}}& \qw & \qw & \cgate{\CChain{\kN}}{CLkcolor} & \qw & \qw
    \end{myqcircuit}
    \quad = \qquad\qquad
    \begin{myqcircuit}[0.1]
        & \cdots &&& \\
        & \qw & \qw & \ctrl{2} & \qw & \ctrl{2} & \qw & \qw \\
        & \qw & \qw & \qw & \qw & \qw & \ctrl{4} & \qw  \\
        & \qw & \qw & \targ{} & \ctrl{2} & \targ{} & \qw & \qw \inputgroupv{1}{5}{1em}{3em}{\hspace{-0.7cm}\ket{\DickeOneU{\n}{\pscoeff}}}\\
        & \cdots &&& \\
        & \qw {/\strut^{\n}}  & \qw & \qw & \cgate{\CChain{\kN}}{CLkcolor} & \qw & \qw & \qw \\
        & \qw {/\strut^{\n}}& \qw & \qw & \qw & \qw & \cgate{\CChain{\kN}}{CLkcolor} & \qw
    \end{myqcircuit}
\end{equation}
On the other hand, we can apply the controlled \copygate{} gate only once by using the following generalization:
\begin{equation}\label{eq:simplification 11}
\begin{myqcircuit}[0.1]
& \cdots && \\
& \qw & \ctrl{6} & \qw & \qw & \qw & \qw & \qw \\
& \qw & \qw & \ctrl{5} & \qw & \qw & \qw & \qw \\
& \qw & \qw & \qw & \ctrl{4} & \qw & \qw & \qw \\
 & \qw & \qw & \qw & \qw & \ctrl{3} & \qw & \qw \\
 & \qw & \qw & \qw & \qw & \qw & \ctrl{2} & \qw \\
 & \cdots && \\
 & \qw & \cgate{\copygate}{ECcolor}&\cgate{\copygate}{ECcolor}&\cgate{\copygate}{ECcolor}&\cgate{\copygate}{ECcolor}&\cgate{\copygate}{ECcolor} & \qw
 \inputgroupv{1}{7}{1em}{4.1em}{\hspace{-2em}\ket{\DickeOneU{\n}{\pscoeff}}}\\ 
\end{myqcircuit}
\; = \qquad\qquad\;
\begin{myqcircuit}[0.1]
& \cdots && \\
& \qw & \ctrl{1} &\qw &\qw &\qw &\qw&\qw &\qw &\qw &\ctrl{1} &\qw\\
& \qw & \targ{} & \ctrl{1} & \qw & \qw & \qw & \qw &\qw &\ctrl{1} &\targ{}  &\qw\\
& \qw & \qw & \targ{} & \ctrl{1} & \qw & \qw & \qw &\ctrl{1}&\targ{} &\qw  &\qw\\
& \qw & \qw & \qw & \targ{} & \ctrl{1} & \qw & \ctrl{1} &\targ{} &\qw&\qw  &\qw\\
& \qw & \qw & \qw & \qw & \targ{} & \ctrl{2} & \targ{} &\qw &\qw &\qw  &\qw\\
& \cdots && \\
& \qw & \qw &\qw &\qw &\qw &\cgate{\copygate}{ECcolor}&\qw &\qw&\qw&\qw  &\qw
\inputgroupv{1}{7}{1em}{4.1em}{\hspace{-2em}\ket{\DickeOneU{\n}{\pscoeff}}}
\end{myqcircuit}
\end{equation}

\subsection{Compression of controlled balanced Dicke states oracles}
\label{subsec: controlled balanced Dicke}
We start by showing that controlling the oracle $\DickeOne{\n}$ defined in \cref{eq_oracle_dicke_one} is equivalent to controlling only the initial $\Xp$ gate, if the initial state is $\ket{0^{\otimes\n}}$, while the subroutine $\staircase{\n}$ does not need to be controlled:
\begin{equation}\label{eq:simplification 1}
\begin{myqcircuit}[0.1]
& \ctrl{1} & \qw \\
\lstick{\ket{0^{\otimes\n}}} & \cgate{\DickeOne{\n}}{DickeOnecolor} & \qw \\
\end{myqcircuit}
\;=\qquad\qquad
\begin{myqcircuit}[0.1]
&& \ctrl{2} & \ctrl{1} & \qw \\
&\lstick{\ket{0^{\otimes\n-1}}} & \qw & \multicgate{1}{\staircase{\n}}{Deltacolor} & \qw \\
&\lstick{\ket{0}} & \gate{\Xp} & \ghost{\staircase{\n}} & \qw  
\end{myqcircuit}
\;=\qquad\qquad
\begin{myqcircuit}[0.5]
&& \ctrl{2} & \qw & \qw \\
&\lstick{\ket{0^{\otimes\n-1}}} & \qw & \multicgate{1}{\staircase{\n}}{Deltacolor} & \qw \\
&\lstick{\ket{0}} & \targ{} & \ghost{\staircase{\n}} & \qw  
\end{myqcircuit}
\end{equation}
This is simply proven by the fact that:
\begin{equation}
    \staircase{\n}\ket{0^{\otimes \n}}=\ket{0^{\otimes \n}} \;.
\end{equation}
This result can be generalized for two controlled $\DickeOne{\n}$ oracles with the same target if the controlling register is a superposition with up to $1$ excitation:
\begin{equation}\label{eq:simplification 6}
\begin{myqcircuit}[0.5]
& \cdots && \\
& \qw & \ctrl{3} &\qw &\qw  \\
& \qw & \qw &\ctrl{2} &\qw \\
& \cdots && \\
\lstick{\ket{0^{\otimes\n-1}}} & \qw & \multicgate{1}{\DickeOne{\n}}{DickeOnecolor} & \multicgate{1}{\DickeOne{\n}}{DickeOnecolor} & \qw \\
\lstick{\ket{0}} & \qw & \ghost{\DickeOne{\n}} & \ghost{\DickeOne{\n}} &\qw \inputgroupv{1}{4}{1em}{2em}{\hspace{-0.8cm}\ket{\DickeOneU{\n}{\pscoeff}}}\\
\end{myqcircuit}
\quad=\qquad\qquad\qquad\quad
\begin{myqcircuit}[0.5]
& \cdots && \\
& \qw & \ctrl{4} &\qw &\qw&\qw  \\
 & \qw & \qw&\ctrl{3} &\qw &\qw\\
 & \cdots && \\
\lstick{\ket{0^{\otimes\n-1}}} & \qw & \qw  &\qw&\multicgate{1}{\staircase{\n}}{Deltacolor} & \qw \\
\lstick{\ket{0}} & \qw & \targ{}&\targ{} & \ghost{\staircase{\n}} & \qw\inputgroupv{1}{4}{1em}{2em}{\hspace{-0.8cm}\ket{\DickeOneU{\n}{\pscoeff}}}
\end{myqcircuit}
\end{equation}
Moreover, given the recursive property from \cref{fig:staircase} and the fact that $\subr{\theta_\n}\ket{00} = \ket{00}$, we get:
\begin{equation}\label{eq:simplification 7}
\begin{myqcircuit}[-0.3]
& \cdots && \\
& \qw & \ctrl{3} &\qw &\qw \\
& \qw & \qw &\ctrl{2} &\qw \\
& \cdots && \\
\lstick{\ket{0^{\otimes\n-1}}} & \qw & \multicgate{1}{\DickeOne{\n}}{DickeOnecolor} & \cgate{\DickeOne{\n-1}}{DickeOnecolor} &\qw \\
\lstick{\ket{0}} & \qw & \ghost{\DickeOne{\n}} & \qw & \qw \inputgroupv{1}{4}{1em}{2em}{\hspace{-2em}\ket{\DickeOneU{\n}{\pscoeff}}}\\
\end{myqcircuit}
\quad = \qquad \qquad \qquad \quad
\begin{myqcircuit}[0.5]
& \cdots && \\
& \qw& \ctrl{5} &\qw &\qw &\qw&\qw \\
& \qw&\qw&\qw &\ctrl{3} &\qw &\qw\\
& \cdots && \\
\lstick{\ket{0^{\otimes\n-2}}} & \qw&\qw  &\qw &\qw &\multicgate{1}{\staircase{\n-1}}{Deltacolor} & \qw \\
\lstick{\ket{0}} & \qw&\qw &\multicgate{1}{\subr{\theta_\n}}{Gammacolor} & \targ{}& \ghost{\staircase{\n-1}} & \qw  \\
\lstick{\ket{0}} & \qw& \targ{} &\ghost{\subr{\theta_\n}} & \qw &\qw &\qw \inputgroupv{1}{4}{1em}{2em}{\hspace{-2em}\ket{\DickeOneU{\n}{\pscoeff}}}\\
\end{myqcircuit}
\end{equation}

\subsection{Compression of controlled unbalanced Dicke states oracles}
\label{subsec: controlled unbalanced Dicke}
In the case of unbalanced Dicke states, we cannot make use of \cref{eq:simplification 6} since the angles of the subroutines $\subr{\theta}$ generally differ.
However, we can leverage the implementation of $\subr{\theta}$ from \cref{section: proof of 2 cnot} to compress every gate that doesn't depend on the encoded coefficients $\pscoeff$.

More specifically, we have:
\begin{equation}
    \begin{myqcircuit}[0.5]
        & \cdots && \\
         & \qw &\ctrl{4} &\ctrl{3} &\qw &\qw &\qw\\
         & \qw &\qw &\qw &\ctrl{3} &\ctrl{2} &\qw \inputgroupv{1}{4}{1em}{2em}{\hspace{-2em}\ket{\DickeOneU{\n}{\pscoeff}}} \\
         & \cdots && \\
         & \qw & \qw &\multicgate{1}{\subr{{\color{mydarkred}\theta_1}}}{Gammacolor} &\qw &\multicgate{1}{\subr{{\color{mypurple}\theta_2}}}{Gammacolor} &\qw\\
         & \qw & \targ &\ghost{\subr{\theta_1}} &\targ &\ghost{\subr{\theta_2}} &\qw 
    \end{myqcircuit}
\end{equation}
and by noticing that:
\begin{equation}
    \begin{myqcircuit}[0.4]
    \lstick{\ket{0}}&\gate{S} &\gate{H} &\ctrl{1}  &\gate{H} &\qw &\ctrl{1} &\gate{S} &\gate{H} &\qw \\
    \lstick{\ket{0}}&\qw&\qw &\targ &\gate{H} &\gate{\conjugate{S}} &\targ &\qw &\qw &\qw 
    \end{myqcircuit}
    \; = \qquad
    \begin{myqcircuit}[0.4]
        \lstick{\ket{0}} & \gate{\Ip} & \qw \\
        \lstick{\ket{0}} & \gate{\Ip} & \qw 
    \end{myqcircuit}
\end{equation}
we can combine them using the decomposition given in \cref{eq: 2cnot subr} as follows:
\begin{equation} \label{eq:compression_gamma_spin_glass}
    \begin{myqcircuit}[-0.2]
        & \cdots && \\
        &\qw &\ctrl{4} &\qw & \qw &\qw &\ctrl{3} &\qw &\qw &\ctrl{4} &\qw &\qw &\qw &\qw &\qw &\qw &\qw \\
        &\qw &\qw &\ctrl{3} &\qw &\qw &\qw &\ctrl{2} &\qw&\qw&\ctrl{3} &\qw&\qw&\qw&\qw&\qw&\qw \inputgroupv{1}{4}{1em}{2.4em}{\hspace{-2em}\ket{\DickeOneU{\n}{\pscoeff}}}\\
        & \cdots && \\
        &\qw &\qw &\qw &\gate{S} &\gate{H} &\cgate{R_z(\frac\pi2-\frac{{\color{mydarkred}\theta_1}}{2})}{Rzcolor} &\cgate{R_z(\frac\pi2-\frac{{\color{mypurple}\theta_2}}{2})}{Rzcolor} &\ctrl{1} &\qw &\qw &\gate{H} &\qw &\ctrl{1} &\gate{S} &\gate{H} &\qw \\
        &\qw &\targ &\targ &\qw&\qw&\qw&\qw &\targ &\cgate{R_z(\frac{{\color{mydarkred}\theta_1}}{2}-\frac{\pi}{2})}{Rzcolor} &\cgate{R_z(\frac{{\color{mypurple}\theta_2}}{2}-\frac{\pi}{2})}{Rzcolor} &\gate{H} &\gate{\conjugate{S}} &\targ &\qw &\qw &\qw 
    \end{myqcircuit}
\end{equation}

\subsection{Compression of \PR{} for the Heisenberg Model}
\label{subsec: compression Heisenberg}
In this section, we present a step-by-step derivation of the optimized circuit shown in \cref{fig:compact_heisenberg}, starting from the original version in \cref{fig:heisenberg_prep}. The goal is to significantly reduce the number of controlled operations.

We begin by decomposing the Dicke state oracles $\DickeTwoNN{\n}$, $\DickeOneD{\n}$, and $\DickeTwoNND{\n}$ using the definitions provided in \cref{section:dicke_two_kN} and \cref{section:dicke_double}. Each oracle is split into two parts: the preparation of the state $\ket{\DickeOne{\n}}$, and either a \CChain{1}, a \copygate{}, or both, depending on the specific oracle.
The resulting circuit looks then like:
\begin{equation}\label{fig:first decomposition}
    \begin{myqcircuit}[0.1]
\lstick{\ket{0}} & \multicgate{5}{\subPR}{subPRcolor} & \ctrl{6} & \qw & \qw & \qw & \qw & \qw & \qw & \qw & \qw & \qw & \qw & \ustick{\textcolor{mygrey}{\gx}}\qw & \\
\lstick{\ket{0}} & \ghost{\subPR} & \qw & \ctrl{6} & \qw & \qw & \qw & \qw & \qw & \qw & \qw & \qw & \qw & \ustick{\textcolor{mygrey}{\gz}}\qw & \\
\lstick{\ket{0}} & \ghost{\subPR} & \qw & \qw & \ctrl{4} & \ctrl{4} & \qw & \qw & \qw & \qw & \qw & \qw & \qw & \ustick{\textcolor{mygrey}{\gy}}\qw & \\
\lstick{\ket{0}} & \ghost{\subPR} & \qw & \qw & \qw & \qw & \ctrl{3} & \ctrl{3} & \qw & \qw & \qw & \qw & \qw & \ustick{\textcolor{mygrey}{\Jx}}\qw & \\
\lstick{\ket{0}} & \ghost{\subPR} & \qw & \qw & \qw & \qw & \qw & \qw & \ctrl{3} & \ctrl{3} & \qw & \qw & \qw & \ustick{\textcolor{mygrey}{\Jz}}\qw & \\
\lstick{\ket{0}} & \ghost{\subPR} & \qw & \qw & \qw & \qw & \qw & \qw & \qw & \qw & \ctrl{1} & \ctrl{1} & \ctrl{1} & \ustick{\textcolor{mygrey}{\Jy}}\qw &\\
\lstick{\ket{0^{\otimes \n}}} & \qw & \cgate{\DickeOne{\n}}{DickeOnecolor} & \qw & \cgate{\DickeOne{\n}}{DickeOnecolor} &\multicgate{1}{\copygate}{ECcolor}& \cgate{\DickeOne{\n-1}}{DickeOnecolor} &\cgate{\CChain{1}}{CLkcolor} &\qw &\qw & \cgate{\DickeOne{\n-1}}{DickeOnecolor} &\cgate{\CChain{1}}{CLkcolor}&\multicgate{1}{\copygate}{ECcolor} &\qw &\\
\lstick{\ket{0^{\otimes \n}}} & \qw & \qw & \cgate{\DickeOne{\n}}{DickeOnecolor} & \qw &\ghost{\copygate} & \qw & \qw & \cgate{\DickeOne{\n-1}}{DickeOnecolor} &\cgate{\CChain{1}}{CLkcolor} & \qw &\qw &\ghost{\copygate} & \qw & \\
\end{myqcircuit}

\end{equation}
Since the $\ket{\subPR}$ state on the six additional ancillae is an unbalanced Dicke state with a single excitation, the gates in \cref{fig:first decomposition} can be rearranged as follows:
\begin{equation}\label{fig: gates move}
    \begin{myqcircuit}[0.1]
\lstick{\ket{0}} & \multicgate{5}{\subPR}{subPRcolor} & \ctrl{6} & \qw & \qw & \qw & \qw & \qw & \qw & \qw & \qw & \qw & \qw & \ustick{\textcolor{mygrey}{\gx}}\qw & \\
\lstick{\ket{0}} & \ghost{\subPR} & \qw & \ctrl{6} & \qw & \qw & \qw & \qw & \qw & \qw & \qw & \qw & \qw & \ustick{\textcolor{mygrey}{\gz}}\qw & \\
\lstick{\ket{0}} & \ghost{\subPR} & \qw & \qw & \ctrl{4} & \qw & \qw & \qw & \qw & \qw & \qw & \ctrl{4} & \qw & \ustick{\textcolor{mygrey}{\gy}}\qw & \\
\lstick{\ket{0}} & \ghost{\subPR} & \qw & \qw & \qw & \ctrl{3} & \qw & \qw & \ctrl{3} & \qw & \qw & \qw & \qw & \ustick{\textcolor{mygrey}{\Jx}}\qw & \\
\lstick{\ket{0}} & \ghost{\subPR} & \qw & \qw & \qw & \qw & \ctrl{3} & \qw & \qw & \ctrl{3} & \qw & \qw & \qw & \ustick{\textcolor{mygrey}{\Jz}}\qw & \\
\lstick{\ket{0}} & \ghost{\subPR} & \qw & \qw & \qw & \qw & \qw & \ctrl{1} & \qw & \qw & \ctrl{1} & \qw & \ctrl{1} & \ustick{\textcolor{mygrey}{\Jy}}\qw & \\
\lstick{\ket{0^{\otimes \n}}} & \qw &\cgate{\DickeOne{\n}}{DickeOnecolor} &\qw &\cgate{\DickeOne{\n}}{DickeOnecolor} & \cgate{\DickeOne{\n-1}}{DickeOnecolor} &\qw & \cgate{\DickeOne{\n-1}}{DickeOnecolor} & \cgate{\CChain{1}}{CLkcolor}&\qw &\cgate{\CChain{1}}{CLkcolor} &\multicgate{1}{\copygate}{ECcolor}&\multicgate{1}{\copygate}{ECcolor} & \qw \\
\lstick{\ket{0^{\otimes \n}}} &\qw &\qw &\cgate{\DickeOne{\n}}{DickeOnecolor} &\qw &\qw &\cgate{\DickeOne{\n-1}}{DickeOnecolor}&\qw &\qw &\cgate{\CChain{1}}{CLkcolor} &\qw &\ghost{\copygate} &\ghost{\copygate} &\qw
\end{myqcircuit}

\end{equation}
In this new configuration, we can apply the compression techniques from \cref{eq:compression_CLk} and \cref{eq:simplification 11} to reduce the number of \CChain{1} and \copygate{} operations, respectively.

Furthermore, the portion of the circuit responsible for preparing balanced Dicke states can be simplified by combining the identities in \cref{eq:simplification 6} and \cref{eq:simplification 7}:
\begin{equation}\label{eq:simplification 8}
\begin{myqcircuit}[0.1]
& \ctrl{6} & \qw & \qw & \qw & \qw & \qw & \ustick{\textcolor{mygrey}{\gx}}\qw \\
& \qw & \ctrl{7} & \qw & \qw & \qw & \qw & \ustick{\textcolor{mygrey}{\gz}}\qw \\
 & \qw & \qw & \ctrl{4} & \qw & \qw & \qw & \ustick{\textcolor{mygrey}{\gy}}\qw \\
& \qw & \qw & \qw & \ctrl{3} & \qw & \qw & \ustick{\textcolor{mygrey}{\Jx}}\qw \\
 & \qw & \qw & \qw & \qw & \ctrl{4} & \qw & \ustick{\textcolor{mygrey}{\Jz}}\qw \\
& \qw & \qw & \qw & \qw & \qw & \ctrl{1} & \ustick{\textcolor{mygrey}{\Jy}}\qw \inputgroupv{1}{6}{1em}{4.6em}{\hspace{-2em}\ket{\DickeOneU{6}{\pscoeff}}} \\
\lstick{\ket{0^{\otimes \n-1}}} &\multicgate{1}{\DickeOne{\n}}{DickeOnecolor} &\qw &\multicgate{1}{\DickeOne{\n}}{DickeOnecolor} & \cgate{\DickeOne{\n-1}}{DickeOnecolor} &\qw & \cgate{\DickeOne{\n-1}}{DickeOnecolor} & \qw \\
\lstick{\ket{0}}& \ghost{\DickeOne{\n}} & \qw & \ghost{\DickeOne{\n}} & \qw & \qw & \qw & \qw \\
\lstick{\ket{0^{\otimes \n-1}}} &\qw &\multicgate{1}{\DickeOne{\n}}{DickeOnecolor} &\qw &\qw &\cgate{\DickeOne{\n-1}}{DickeOnecolor}&\qw &\qw \\
\lstick{\ket{0}} & \qw & \ghost{\DickeOne{\n}} & \qw & \qw & \qw & \qw & \qw
\end{myqcircuit}
\raisebox{0.5cm}{$\;\quad = \qquad\qquad\;$}
\begin{myqcircuit}[0.62]
&\ctrl{8} &\qw &\qw &\qw &\qw&\qw &\qw &\qw & \ustick{\textcolor{mygrey}{\gx}}\qw \\
& \qw &\ctrl{10} &\qw&\qw &\qw&\qw &\qw &\qw & \ustick{\textcolor{mygrey}{\gz}}\qw \\
& \qw &\qw &\ctrl{6}&\qw &\qw&\qw &\qw&\qw & \ustick{\textcolor{mygrey}{\gy}}\qw \\
&\qw&\qw &\qw &\qw &\ctrl{4}&\qw &\qw &\qw & \ustick{\textcolor{mygrey}{\Jx}}\qw\\
& \qw &\qw &\qw&\qw &\qw&\ctrl{6} &\qw &\qw & \ustick{\textcolor{mygrey}{\Jz}}\qw\\
& \qw&\qw &\qw&\qw &\qw&\qw&\ctrl{2} &\qw & \ustick{\textcolor{mygrey}{\Jy}}\qw \inputgroupv{1}{6}{1em}{3.8em}{\hspace{-2em}\ket{\DickeOneU{6}{\pscoeff}}}\\
\lstick{\ket{0^{\otimes \n-2}}} &\qw&\qw &\qw&\qw &\qw &\qw &\qw &\multicgate{1}{\staircase{\n-1}}{Deltacolor}&\qw \\
\lstick{\ket{0}}&\qw&\qw &\qw &\multicgate{1}{\subr{\theta_{\n}}}{Gammacolor} &\targ{} &\qw &\targ{} &\ghost{\staircase{\n-1}}&\qw \\
\lstick{\ket{0}}&\targ{}&\qw &\targ{} &\ghost{\subr{\theta_{\n}}}&\qw &\qw &\qw &\qw&\qw\\
\lstick{\ket{0^{\otimes \n-2}}} &\qw&\qw&\qw &\qw&\qw &\qw &\qw &\multicgate{1}{\staircase{\n-1}}{Deltacolor}&\qw \\
\lstick{\ket{0}} &\qw&\qw&\qw &\multicgate{1}{\subr{\theta_{\n}}}{Gammacolor}&\qw &\targ{}&\qw &\ghost{\staircase{\n-1}}&\qw\\
\lstick{\ket{0}} &\qw &\targ{}&\qw &\ghost{\subr{\theta_{\n}}}&\qw &\qw &\qw &\qw &\qw
\end{myqcircuit}
\end{equation}
This sequence of optimizations leads to the final, compressed implementation of the \PR{} circuit for the Heisenberg model shown in \cref{fig:compact_heisenberg}.

\subsection{Compression of \PR{} for the Spin Glass model}
\label{subsec: compression Spin Glass}
The simplification of the Spin Glass model follows a similar approach to the Heisenberg case.
For each fixed value of $\kN$, we begin by separating the preparation of the unbalanced Dicke state with $\excitations = 1$ from the subsequent \CChain{\kN} and \copygate{} operations:
\begin{gather}
\text{$\begin{myqcircuit}[-0.2]
& \cdots && \\
& \qw &\ctrl{4} & \qw & \qw & \ustick{\textcolor{mygrey}{\diagnorm{\xsub}{\kN}}}\qw & \\
& \qw &\qw & \ctrl{4} & \qw & \ustick{\textcolor{mygrey}{\diagnorm{\zsub}{\kN}}}\qw & \\
& \qw &\qw & \qw & \ctrl{2} & \ustick{\textcolor{mygrey}{\diagnorm{\ysub}{\kN}}}\qw  \inputgroupv{1}{5}{1em}{3.5em}{\hspace{-2em}\ket{\DickeOneU{\n}{\pscoeff}}} \nonumber\\
& \cdots && \\
& \qw &\cgate{\DickeTwokNU{\n}{\kN}{\coefmatnormnum{\xsub}{\kN}}}{DickeTwokNUcolor} & \qw & \multicgate{1}{\DickeTwokNUD{\n}{\kN}{\coefmatnormnum{\ysub}{\kN}}}{DickeTwokNUDcolor} & \qw & \\
& \qw &\qw & \cgate{\DickeTwokNU{\n}{\kN}{\coefmatnormnum{\zsub}{\kN}}}{DickeTwokNUcolor} & \ghost{\DickeTwokNUD{\n}{\kN}{\coefmatnormnum{\ysub}{\kN}}} & \qw & 
\end{myqcircuit}$}\\
\hspace{-2cm}= \qquad\qquad
\text{$\begin{myqcircuit}[-0.2]
    & \cdots && \\
    &\qw&\ctrl{4} &\ctrl{4} &\qw&\qw&\qw&\qw&\qw& \ustick{\textcolor{mygrey}{\diagnorm{\xsub}{\kN}}}\qw\\
    &\qw&\qw&\qw&\ctrl{4} &\ctrl{4} &\qw&\qw&\qw&\ustick{\textcolor{mygrey}{\diagnorm{\zsub}{\kN}}}\qw\\
    &\qw&\qw&\qw&\qw &\qw &\ctrl{2}&\ctrl{2} &\ctrl{2}&\ustick{\textcolor{mygrey}{\diagnorm{\ysub}{\kN}}}\qw \inputgroupv{1}{5}{1em}{3.5em}{\hspace{-2em}\ket{\DickeOneU{\n}{\pscoeff}}}\\
    & \cdots && \\
    &\qw&\cgate{\DickeOneU{\n-\kN}{\coefmatnormnum{\xsub}{\kN}}}{DickeOneUcolor} &\cgate{\CChain{\kN}}{CLkcolor} &\qw &\qw &\cgate{\DickeOneU{\n-\kN}{\coefmatnormnum{\ysub}{\kN}}}{DickeOneUcolor} &\cgate{\CChain{\kN}}{CLkcolor} & \multicgate{1}{\copygate{}}{ECcolor} & \qw \\
    &\qw&\qw &\qw &\cgate{\DickeOneU{\n-\kN}{\coefmatnormnum{\zsub}{\kN}}}{DickeOneUcolor} &\cgate{\CChain{\kN}}{CLkcolor} &\qw &\qw &\ghost{\copygate{}} & \qw
\end{myqcircuit}$}
\end{gather}
As shown in \cref{fig: gates move}, and since the \subPR{} subroutine prepares the extra ancillae in an unbalanced Dicke state with $\excitations=1$, we can group together all the Dicke oracles first and move all the \CChain{\kN} and \copygate{} at the end of the circuit.
Then, for each value of $\kN$, we reduce the number of \CChain{\kN} subroutines from three to two by applying the state identity in \cref{eq:compression_CLk}.
In addition, the entire circuit requires only a single \copygate{} operation, as shown in \cref{eq:simplification 11}.

\begin{figure*}
    \centering
    \subfloat[]{\label{fig:spinglass_complete_1}\includegraphics[scale=0.9]{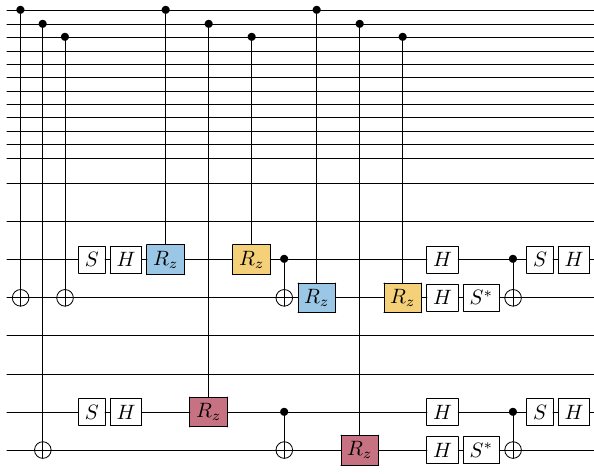}} \\
    \subfloat[]{\label{fig:spinglass_complete_2}\includegraphics[scale=0.9]{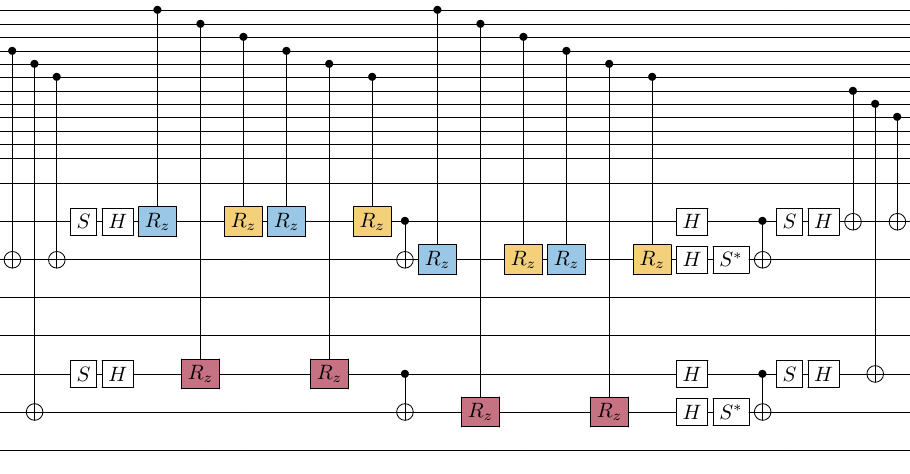}}\\
    \subfloat[]{\label{fig:spinglass_complete_3}\includegraphics[scale=0.9]{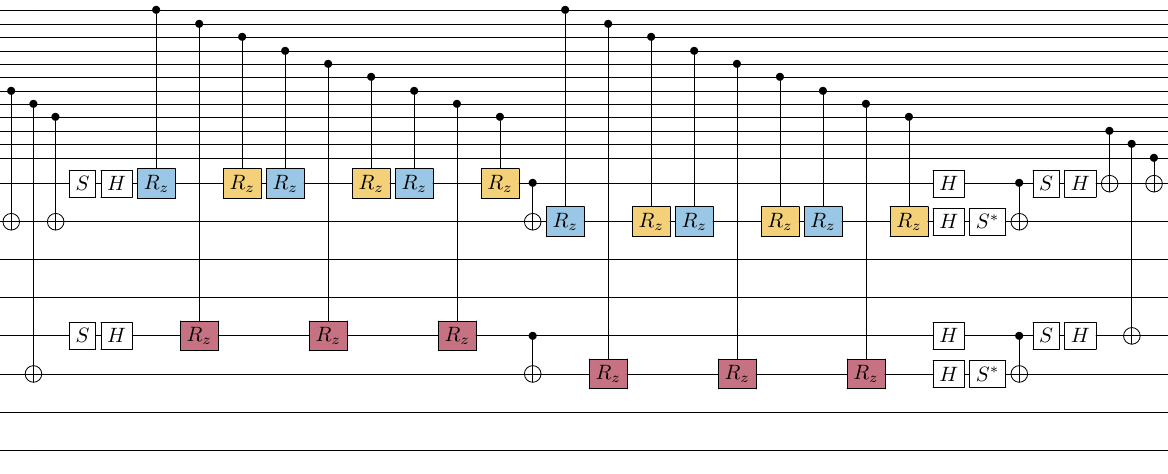}}
    \caption{Example with $\n=4$ of the compression of the oracles $\DickeOneU{\n-\kN}{\coefmatnormnum{}{\kN}}$ for the spin glass model.}
\label{fig:compression_dicke_one_unbalanced_spin_glass}
\end{figure*}

Finally, for what concerns the $\DickeOneU{\n-\kN}{\coefmatnormnum{}{\kN}}$ oracles, we make use of the simplifications shown in \cref{subsec: controlled unbalanced Dicke}. An example for $\n=4$ is shown in \cref{fig:compression_dicke_one_unbalanced_spin_glass}.

\section{General two-body interactions}
\label{app: generalization}

In this section, we show how to further generalize the \PR{} constructions presented in \cref{section:applications} by considering the two-body operators $\Xp_{\ell}\Yp_{\ell+\kN}$, $\Yp_{\ell}\Xp_{\ell+\kN}$, $\Xp_{\ell}\Zp_{\ell+\kN}$, $\Zp_{\ell}\Yp_{\ell+\kN}$, $\Zp_{\ell}\Xp_{\ell+\kN}$ and $\Yp_{\ell}\Zp_{\ell+\kN}$.

\begin{figure*}
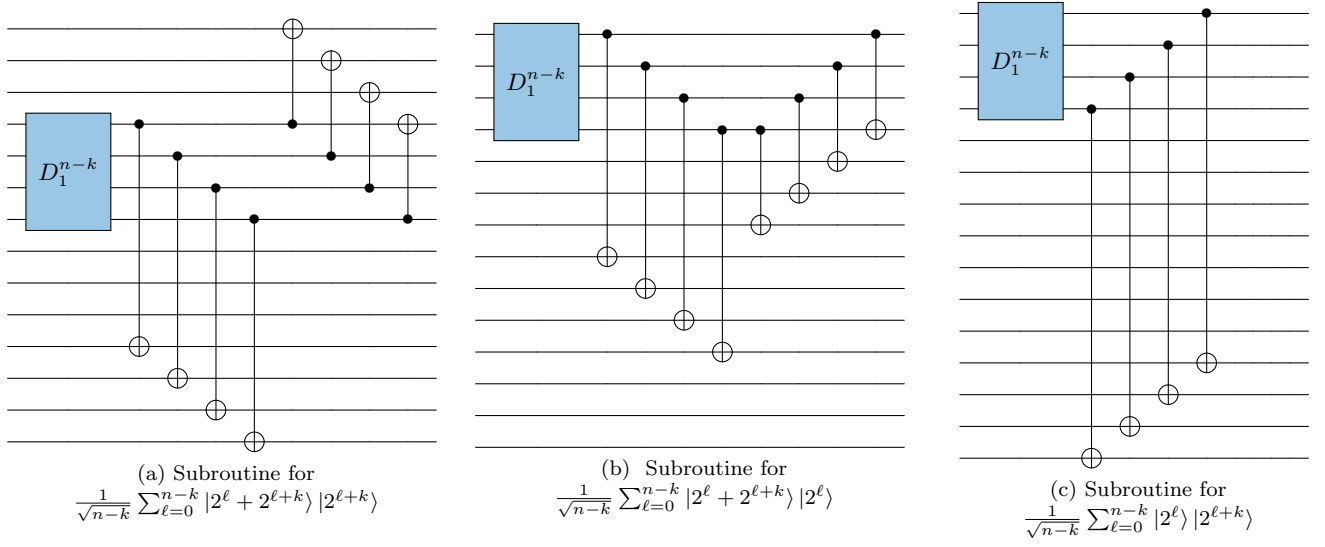

    \centering
    ~\hfill
    \subfloat[Subroutine for $\frac1{\sqrt{\n-\kN}}\sum_{\ell =0}^{\n-\kN}\ket{2^{\ell}+2^{\ell+\kN}}\ket{2^{\ell+\kN}}$]{\label{fig:generalization_example_xy}\include{figures/appendix_E_1}}
    \hfill
    \subfloat[ Subroutine for $\frac1{\sqrt{\n-\kN}}\sum_{\ell =0}^{\n-\kN}\ket{2^{\ell}+2^{\ell+\kN}}\ket{2^{\ell}}$]{\label{fig:generalization_example_yz}\include{figures/appendix_E_2}}
    \hfill~
    \subfloat[Subroutine for $\frac1{\sqrt{\n-\kN}}\sum_{\ell =0}^{\n-\kN}\ket{2^{\ell}}\ket{2^{\ell+\kN}}$]{\label{fig:generalization_example_xz}\include{figures/appendix_E_3}}
    \caption{Circuits for prepare states for generalize two body interaction with $\n = 7$ and $\kN = 3$}
    \label{fig:generalization_example}
\end{figure*}

First of all, by generalizing the operator-to-state mapping shown in \cref{tab:ancilla_operators} we get:
\begin{align}
    \Xp_{\ell}\Yp_{\ell+\kN} &\rightarrow \ket{2^\ell+2^{\ell+\kN}}\ket{2^{\ell+\kN}} \\
    \Yp_{\ell}\Xp_{\ell+\kN} &\rightarrow \ket{2^\ell+2^{\ell+\kN}}\ket{2^\ell} \\
    \Xp_{\ell}\Zp_{\ell+\kN} &\rightarrow \ket{2^\ell}\ket{2^{\ell+\kN}} \\
    \Zp_{\ell}\Yp_{\ell+\kN} &\rightarrow \ket{2^{\ell+\kN}}\ket{2^\ell+2^{\ell+\kN}}\\
    \Yp_{\ell}\Zp_{\ell+\kN} &\rightarrow \ket{2^\ell}\ket{2^\ell+2^{\ell+\kN}} \\
    \Zp_{\ell}\Xp_{\ell+\kN} &\rightarrow \ket{2^{\ell+\kN}}\ket{2^\ell} \;.
\end{align}
The balanced sum of operators $\frac1{\sqrt{\n-\kN}}\sum_{\ell =0}^{\n-\kN-1}\Xp_{\ell}\Yp_{\ell+\kN}$ corresponds then to the state $\frac1{\sqrt{\n-\kN}}\sum_{\ell =0}^{\n-\kN-1}\ket{2^\ell+2^{\ell+\kN}}\ket{2^{\ell+\kN}}$, which can be implemented by the following circuit:
\begin{equation}
    \begin{myqcircuit}
        &\qw&{/\strut^{\kN}}\qw &\qw &\qw &\multicgate{2}{\copygate{}}{ECcolor} &\multicgate{1}{\CChain{-\kN}}{CLkcolor} &\qw\\
        &\qw&{/\strut^{\n - \kN}}\qw &\qw &\cgate{\DickeOne{\n -\kN}}{DickeOnecolor} &   \ghost{\copygate{}}& \ghost{\CChain{-\kN}} &\qw \\
        &\qw&{/\strut^{\n}}\qw &\qw &\qw &\ghost{\copygate{}} &\qw &\qw
    \end{myqcircuit}
\end{equation}
With \CChain{-\kN} we denote a variant of the \CChain{\kN} gate where the target is applied $\kN$ qubits \textit{above} the control.
An example with $\n=7$ and $\kN=3$ can be seen in \cref{fig:generalization_example_xy}.

The proof is as follows:
\begin{align}
    \ket{0^{\otimes \n}}\ket{0^{\otimes \n}} &\xrightarrow{I_{\kN}\otimes\DickeOne{\n-\kN-1}\otimes I_{\n}}\ket{0^{\otimes \kN}} \frac1{\sqrt{\n-\kN}}\sum_{\ell =0}^{\n-\kN}\ket{2^{\ell}}\ket{0^{\otimes \n}} \nonumber\\
    &\xrightarrow{\copygate{}}\frac1{\sqrt{\n-\kN}} \sum_{\ell =0}^{\n-\kN-1}\ket{0^{\otimes \kN}} \ket{2^{\ell}} \ket{0^{\otimes \kN}}\ket{2^\ell} \nonumber \\
    & = \frac1{\sqrt{\n-\kN}} \sum_{\ell =0}^{\n-\kN-1} \ket{2^{\ell+\kN}} \ket{2^{\ell+\kN}} \nonumber \\
    &\xrightarrow{\CChain{-\kN}} \frac1{\sqrt{\n-\kN}} \sum_{\ell =0}^{\n-\kN-1} \ket{2^{\ell}+2^{\ell+\kN}} \ket{2^{\ell+\kN}}
\end{align}
where we remark the fact that $\ket{0^{\otimes \kN}} \ket{2^{\ell}} = \ket{2^{\kN+\ell}} $.

To implement the state $\sum_{\ell=0}^{n-k_N-1} \ket{2^{\ell+k_N}}\ket{2^\ell + 2^{\ell+k_N}}$, which corresponds to the operator $\sum_{\ell=0}^{n-k_N-1} Z'_{\ell} Y'_{\ell+k_N}$, we can proceed in the $z$-register analogously to how we have previously operated in the $x$-register. This can be achieved by using the subroutine $\copygate{}$, modifying it so that the roles of the control and target qubits are swapped.

The operator $\frac1{\sqrt{\n-\kN}}\sum_{\ell =0}^{\n-\kN-1}\Yp_{\ell}\Xp_{\ell+\kN}$ maps to the state $\frac1{\sqrt{\n-\kN}}\sum_{\ell =0}^{\n-\kN}\ket{2^{\ell}+2^{\ell+\kN}}\ket{2^{\ell}}$ with implementing circuit:
\begin{equation}
    \begin{myqcircuit}
        &\qw&{/\strut^{\n-\kN}}\qw &\qw&\cgate{\DickeOne{\n -\kN}}{DickeOnecolor} &\multicgate{2}{\copygate{}}{ECcolor} &\multicgate{1}{\CChain{\kN}}{CLkcolor} &\qw\\
        &\qw&{/\strut^{\kN}}\qw &\qw&\qw &   \ghost{\copygate{}}& \ghost{\CChain{\kN}} &\qw \\
        &\qw&{/\strut^{\n}}\qw &\qw&\qw &\ghost{\copygate{}} &\qw &\qw
    \end{myqcircuit}
\end{equation}

As before, we can implement the state for the operator $\frac1{\sqrt{\n-\kN}}\sum_{\ell =0}^{\n-\kN-1}\Yp_{\ell}\Zp_{\ell+\kN}$ by applying the subroutine that we used in the $z$-register instead of the $x$-register.

On the other hand, for $\frac1{\sqrt{\n-\kN}}\sum_{\ell =0}^{\n-\kN-1}\Xp_{\ell}\Zp_{\ell+\kN}$ and the corresponding state $\frac1{\sqrt{\n-\kN}}\sum_{\ell =0}^{\n-\kN}\ket{2^{\ell}}\ket{2^{\ell+\kN}}$ we have:
\begin{equation}
\label{eq: circuit ZX}
    \begin{myqcircuit}
        &{/\strut^{\n-\kN}}\qw &\qw&\cgate{\DickeOne{\n -\kN}}{DickeOnecolor} &\multicgate{2}{\CChain{\n+\kN}}{CLkcolor} &\qw\\
        &{/\strut^{\kN}}\qw &\qw&\qw & \ghost{\CChain{\n+\kN}} &\qw \\ 
        &{/\strut^{\n}}\qw &\qw&\qw & \ghost{\CChain{\n+\kN}} &\qw \\ 
    \end{myqcircuit}
\end{equation}
This is trivial since we have that $\ket{2^\ell}\ket{2^{\ell+\kN}} = \ket{2^\ell+2^{\ell+\kN+\n}}$

Once again, for the operator $\frac{1}{\sqrt{n - k_N}} \sum_{\ell = 0}^{n - k_N - 1} Z'_{\ell} X'_{\ell + k_N}$, the circuit has the same structure as \cref{eq: circuit ZX}; the only difference is that the subroutine needs to be applied in the $z$-register.

In \cref{fig:generalization_example}, we illustrate the three circuits for the specific case $\n=7$ and $\kN=3$.
Note that some of the \cnots{} involved in the definition of the \copygate{} can be omitted by assuming the initial state is always $\ket{0^{\otimes 2\n}}$.
Moreover, the circuits described in this section can be extended to the unbalanced case by replacing the subroutine $\DickeOne{\n-\kN}$ with its unbalanced counterpart, as detailed in \cref{section:unbalanced_dicke}.

\end{document}